\documentclass[useAMS,usenatbib]{mn2e}

\usepackage{graphicx}
\usepackage{subfigure}
\usepackage{epsfig}
\usepackage{color}
\usepackage{lmodern}
\title[Morphological Evolution in XMMU J1229+0151]{The Morphological Transformation of Red-Sequence Galaxies in the Distant Cluster XMMU J1229+0151}
\author[P. Cerulo et al.]{\parbox{\textwidth}{P. Cerulo$^{1}$\thanks{E-mail:
pcerulo@astro.swin.edu.au}, W. J. Couch{$^{1,2}$}, C. Lidman$^{2}$, L. Delaye$^{3,6}$, R. Demarco$^{4}$, M. Huertas-Company$^{3}$, S. Mei$^{3}$, R. S\'anchez-Janssen$^{5}$}\vspace{0.4cm}\\
\parbox{\textwidth}{$^{1}$Centre for Astrophysics and Supercomputing, Swinburne University of Technology, PO Box 218, Hawthorn, VIC 3122, Australia\\
$^{2}$Australian Astronomical Observatory, PO Box 915, North Ryde, NSW 1670, Australia\\
$^{3}$GEPI, Paris Observatory, 77 av. Denfert Rochereau, 75014 Paris, France\\
$^{4}$Department of Astronomy, Universidad de Concepcion, Casilla 160-C Concepcion, Chile\\
$^{5}$NRC Herzberg Institute of Astrophysics, 5071 West Saanich Road, Victoria, BC V9E 2E7, Canada \\
$^{6}$CEA-Saclay, DSM/IRFU/SAp, F-91191 Gif-sur-Yvette, France}}
\begin{document}


\pagerange{\pageref{firstpage}--\pageref{lastpage}} \pubyear{2014}

\maketitle

\label{firstpage}

\begin{abstract}
We present the results of a detailed analysis of galaxy properties along the red
sequence in XMMU J1229+0151, an X-ray selected cluster at $z=0.98$ drawn from the HAWK-I Cluster Survey (HCS). Taking
advantage of the broad photometric coverage and the availability of 77
spectra in the cluster field, we fit synthetic spectral energy distributions, and
estimate stellar masses and photometric redshifts, which we use to determine the
cluster membership. We investigate
morphological and structural properties of red sequence galaxies and find that
elliptical galaxies populate the bright end, while S0 galaxies represent the
predominant population at intermediate luminosities, with their fraction
decreasing at fainter magnitudes. 
A comparison with the low-redshift sample of the WINGS cluster survey reveals that at $z\sim1$ the bright end of the red sequence of XMMU J1229+0151 is richer in S0 galaxies. The faint end of the red sequence in XMMUJ1229+0151 appears rich in disc-dominated galaxies, which are rarer in the low redshift comparison sample at the same luminosities. \textcolor{black}{Despite these differences between the morphological composition of the red sequence in XMMUJ1229+0151 and in low redshift samples, we find that to within the uncertainties, no such difference exists in the ratio of luminous to faint galaxies along the red sequence.}
\end{abstract}


\section{Introduction}

Clusters of galaxies are the most massive virialised large scale
structures in the universe and, thanks to the broad range of densities available
in these systems, they can be used as laboratories for the study of the
environmental drivers of galaxy evolution \citep{De_Lucia_2007}. According to the hierarchical scenario of structure formation, galaxy clusters form after the collapse of the highest density peaks in the primordial matter
distribution, accreting other smaller dark matter haloes at later times. By the present day, they have built up into systems whose total mass can reach up to 10$^{15}$ $M_\odot$, with characteristic sizes of a couple of Mpc.

Up to $z\sim1.5$, red and passive elliptical and S0 galaxies are the predominant population in cluster cores, whereas blue and star-forming spiral and irregular galaxies dominate the outskirts (see \citealt{Dressler_1980, Dressler_1997, Postman_2005, Hilton_2009, Muzzin_2012, Mei_2012}). \textcolor{black}{However, the fraction of blue star forming galaxies in clusters increases with redshift \citep{Butcher_1978} and above $z \sim 1.5$ there is evidence of star formation in cluster cores \citep{Hilton_2010, Tran_2010, Hatch_2011, Hayashi_2012} These results suggest that most of the processes that led to the establishment of the star-formation- and morphology-local density relations, as they are observed in local clusters, were active in the interval $1.0 < z < 1.5$.}

A large number of mechanisms have been proposed to both trigger star formation and then quench it. Galaxy-galaxy merging within groups that are falling into the cluster for the first time, galaxy-galaxy harassment within the clusters as galaxies pass each other at high velocities, and collisional compression and ram pressure stripping of gas in galaxies by the hot intracluster medium are mechanisms that are all thought to be at play in the dense cluster environment \citep{Gunn_1972, Lavery_1988, Moore_1996, Bekki_1999, Bekki_2003}. However, from the current observations, it is not clear which of these processes is driving galaxy evolution in clusters. Furthermore, clusters are highly heterogeneous systems, and core and outskirts constitute 
different environments, with different global physical properties. The interactions of the galaxies with their surroundings may therefore change substantially according to their location within the cluster.

Regardless of the environment in which they reside, galaxies have a bimodal colour distribution, such that the colour-magnitude diagram of a galaxy population is characterised by a narrow sequence of red quiescent objects and a diffuse cloud of blue star-forming galaxies. As galaxies finish to form stars, depleting their gas reservoirs, they migrate from the blue cloud to the red sequence. Thus, in principle, the evolution of a galaxy population can be investigated by looking at the gradual build-up of the red sequence with redshift. However, bursts of star formation resulting from merger events between a quiescent and a star-forming galaxy may push the galaxy back to the blue cloud (see e. g.\ \citealt{Faber_2007}). 

\cite{Kodama_1997} explained the red sequence as a mass-metallicity relation, the scatter about the best-fit straight line being driven by age differences among galaxies. However, \cite{Gallazzi_2006} demonstrated that metallicity contributes to the scatter 
too and that the age-induced scatter is anti-correlated with stellar mass.  As the local galaxy density increases, the red sequence becomes more pronounced with respect to the blue cloud (see e. g.:\ \citealt{Hogg_2004}) and galaxy clusters are characterised by a tight and well defined red sequence, which can be used to estimate the cluster redshift when no spectroscopic information is available \citep{Gladders_2005, Andreon_2011}.

Between $z=1$ and $z=0$, the number of red sequence galaxies fainter than $M_V = -20.0$ is found to increase, approximately halving the relative ratio between luminous (i.e. $M_V < -20.0$) and faint galaxies ($M_V \geq -20.0$) (see e. g.\ \citealt{Tanaka_2005, De_Lucia_2007, Gilbank_2008, Capozzi_2010, Bildfell_2012, Lemaux_2012}). This deficit of galaxies at the faint end of the red sequence has been detected in clusters up to redshift $z=1.62$
\citep{Rudnick_2012}. However, \cite{Andreon_2008} studied the trend of the luminous to faint ratio in a sample of 28 galaxy clusters at $0.0 < z < 1.3$, finding no deficit. The existence of a deficit of galaxies at the faint end of the red sequence supports the notion of a build-up at low masses. Low-mass galaxies quench their star formation at later times, with respect to higher-mass systems, and therefore they join the red sequence at later times. This property, which is known as {\itshape{downsizing}} \citep{Cowie_1996}, is observed also in the field up to $z \sim 2$ (see e.g.\ \citealt{Tanaka_2005, Ilbert_2010}). 

\cite{Muzzin_2012} investigated the separate effects of mass and environment on the evolution of galaxies in a sample of clusters at $0.8 < z < 1.4$ from the Gemini Cluster Astrophysics Spectroscopic Survey (GCLASS). They observed a decrease of the fraction of star-forming galaxies at increasing galaxy stellar masses and local densities. They found that, at fixed environment (i.e. cluster core and outskirts, and field), the star formation rate (SFR), specific star formation rate (SSFR) and the amplitude of the 4000 \AA\ break ($D_n(4000)$) were all correlated with stellar mass. However, at fixed stellar mass, the same quantities were not found to correlate with the environment. They concluded that stellar mass is the primary driver for the quenching of star formation, regardless of the environment, while the environment acts on the global galaxy population quenching the star formation rate over a relatively short period of time, regardless of galaxy stellar mass. In other words, the effect of the environment 
is 
to accelerate 
star-formation quenching leaving the correlations with mass of SFR, SSFR and $D_n(4000)$ unchanged. The conclusions of \cite{Muzzin_2012} extended to higher galaxy densities what had previously been observed in the field up to $z = 2$ by \cite{Peng_2010, Peng_2012} and  \cite{Quadri_2012}. 

While \cite{Peng_2010} and \cite{Muzzin_2012} suggest that environment plays a role in accelerating the shutting down of star-formation, neither explore the mechanisms by  which environment does this. \cite{Demarco_2010} investigated the properties of red sequence members in the cluster RX J0152.7-1357, at $z = 0.84$, finding that early-type galaxies at the faint end of the red sequence are systematically younger than galaxies at the bright end of the red sequence and are preferentially located in the cluster outskirts. They concluded that star formation in these galaxies had
recently been quenched either through ram pressure stripping of gas by the hot intracluster medium, or rapid exhaustion of gas reservoirs from an earlier epoch of star formation caused by galaxy-galaxy merging. These results support once again the notion of a build-up of the red sequence at low masses, suggesting that the evolution of cluster galaxies fits in the downsizing scenario.

We extend the work of \cite{Demarco_2010} by making a comprehensive study of the morphological and structural properties of red sequence galaxies in the cluster XMMU J1229+0151 \citep[hereafter XMM1229,][ Fig. 1]{Santos_2009}, at $z=0.98$. The cluster is part of the HAWK-I Cluster Survey (HCS, \citealt{Lidman_2013}), comprising a sample of 9 galaxy clusters at $0.8 < z < 1.5$. At the low redshift end in the HCS sample there is the cluster RX J0152.7-1357, whose galaxy population was studied in detail by \cite{Demarco_2010}. XMM1229 has imaging and spectroscopic data from several space and ground based facilities and its X-ray and dark matter properties were studied by \cite{Santos_2009} and \cite{Jee_2011}, respectively. Furthermore, \cite{Santos_2009} also analysed the properties of the spectroscopically confirmed members of this cluster.  \textcolor{black}{The present work extends the analysis of \cite{Santos_2009} to other cluster members selected with photometric redshifts, whose estimation has been 
possible thanks to the additional imaging data become available after the publication of that work and described in \S 2.} 

The position of the cluster in the redshift range of the HCS and the availability of a considerable quantity of data make XMM1229 suitable to develop a method for the analysis of the red sequence in the HCS clusters. The method developed in this paper will be extended to the other clusters of the HCS sample in order to study the build-up of the red sequence in galaxy clusters at $0.8 < z < 1.5$. In particular, our interest will be focused on three points: the characteristics of the red sequence itself (its shape, slope, and scatter), the ratio between luminous and faint galaxies, and the morphological properties of galaxies along the red sequence. The availability of the Wide Field Nearby Galaxy-clusters Survey (WINGS, \citealt{Fasano_2006}) and the MORPHS survey \citep{Smail_1997} allow us to build comparison samples of clusters at $z \sim 0$ and $0.3 < z < 0.6$, respectively, which are used to compare the properties of the red sequence members 
in XMM1229 and, in a forthcoming paper, of the entire HCS sample. 

The paper is organised as follows: we describe the observations and data analysis in \S\S 2. and 3. \S 4 presents the results of our study which are discussed in \S 5, while we sumamarise our work and draw our conclusions in \S 6. Throughout the paper we use a $\Lambda$CDM cosmology with $H_0 = 71$ km $\cdot$ s$^{-1}$ $\cdot$ Mpc$^{-1}$, $\Omega_M = 0.27$, and $\Omega_{\Lambda} = 0.73$. All magnitudes are
quoted in the AB system \citep{Oke_1978}, unless stated otherwise.

\begin{figure*}
	\includegraphics[trim=0.0cm 2.5cm 0.0cm 0.0cm, clip, width=16cm]{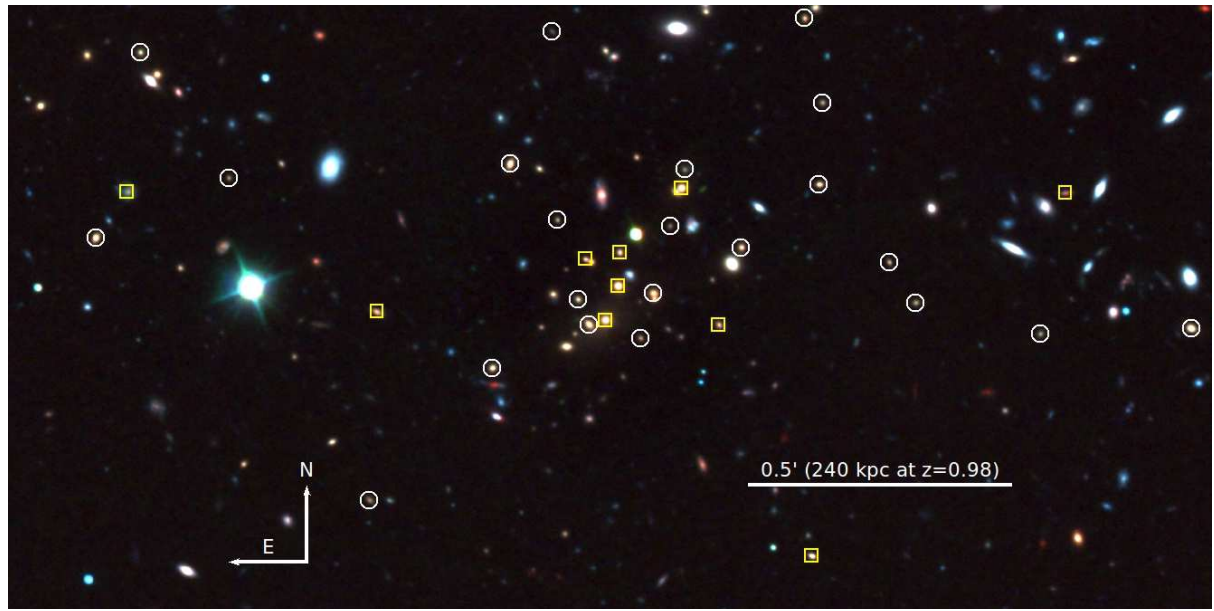}
	\caption{Colour image of XMM1229 obtained by combining the ACS F775W and F850LP images, and the HAWK-I Ks image. White circles are photometrically selected red sequence members and yellow squares are spectroscopically confirmed members (see \S 4.1 and Table 3 for the determination of the cluster membership and a complete list of red sequence galaxies, respectively).}
\end{figure*}

\section{Observations and data reduction} 

XMM1229 was first discovered as an X-ray overdensity in the XMM-Newton Distant
Cluster Project (XDCP) \citep{Fassbender_2011}, a survey that used XMM-Newton telescope data to detect distant galaxy clusters. The cluster has been later targeted by several space- and ground-based telescopes (HST, VLT, NTT), providing us
with a rich dataset covering the spectral region $0.65 < \lambda <
2.2\mu m$ (Fig. 2), as well as spectra for 26 cluster members. With the data having been acquired at different
times and on different telescopes, with strategies varying
with each program, the resulting data set is very diverse. This section describes this rich multi-wavelength data set detailing in particular the reduction methods we have employed to be able to utilise it as a complete ensemble.

\begin{figure}
  \centering
	\includegraphics[width=6.5 cm]{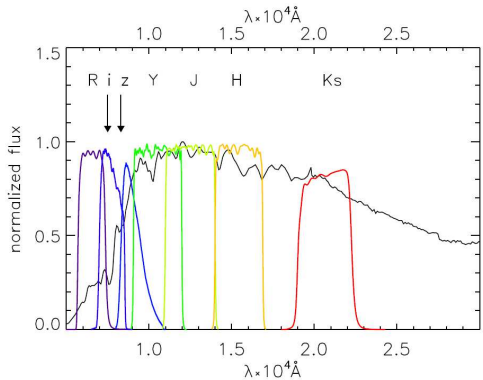}
	\caption{Photometric coverage of the XMM1229 field. From
left to right: R\_SPECIAL (R), F775W (i), F850LP (z), F105W (Y), F125W (J), F160W (H), Ks. The F110W transmission curve is not plotted, as it covers the same spectral range of the F105W and F125W bands. The SofI J band transmission curve is not plotted because it overlaps with F125W. The F775W and F850LP bands used for the colour-magnitude diagram are highlighted by the arrows. The solid black line represents the template SED of an elliptical galaxy from \protect\cite{Coleman_1980} at the redshift of XMM1229.}
\end{figure}

\subsection{HST imaging}

\subsubsection{Advanced Camera for Surveys (ACS)}

XMM1229 was imaged in November 2006 in the F775W ($i_{775}$) and F850LP
($z_{850}$) bands of the Wide Field Channel (WFC) of the Advanced Camera for
Surveys (ACS), on board the Hubble Space Telescope (HST). The observations were part of the Hubble
Space Telescope Cluster Supernova Survey \citep{Dawson_2009}, aimed at the
search of Type Ia supernovae (SNeIa) in cluster galaxies at $0.9<z<1.5$.

We summarise here the survey strategy and the data reduction process for the ACS
observations of XMM1229, referring to \cite{Dawson_2009} and \cite{Suzuki_2012}
for a more detailed description. The HST Cluster Supernova Survey collected
$i_{775}$ and $z_{850}$ observations of 25 X-ray, optically or infrared (IR) detected galaxy
clusters over the redshift range $0.9<z<1.5$. The clusters
were observed in multiple visits and, in each visit, at least one exposure in $i_{775}$
and three or four exposures in $z_{850}$ were collected. The images were calibrated using the standard
calibration pipeline provided by the Space Telescope Science Institute (STScI)
and were registered and stacked using {\ttfamily Multidrizzle} \citep{Fruchter_2002, Koekemoer_2002}, with a final pixel scale of 0.05 $''$/pixel for all the
clusters. XMM1229 was observed during 8 visits, with 9 and 30 exposures
respectively collected for the $i_{775}$ and $z_{850}$ bands. The resulting
image covers 5$'$.1 $\times$ 5$'$.1, with an image quality{\footnote{Throughout this paper we use the FWHM of stars as a  measure of the image quality.}} of $\sim$ 0.09$''$ in both
bands. The details of the ACS observations are reported in Table 1.

\subsubsection{Wide Field Camera 3 (WFC3)}

XMM1229 was imaged in the IR channel of Wide Field Camera 3 (WFC3),
on board HST, as part of the program 12051 (P. I.: S.
Perlmutter), aimed at the calibration of the sensitivities of NICMOS and WFC3
for faint objects. The observations were taken in 2010 May 24 in the F105W (Y),
F110W, F125W (J) and F160W (H) filter bands, following a BOX-MIN dithering
pattern. The images were reduced with {\itshape calwf3}, using the most recent
versions of the calibration frames available for WFC3. We combined 
the reduced images ({\ttfamily{\_flt}} files) with {\ttfamily Multidrizzle}, setting the drop size ({\ttfamily pixfrac} parameter)  to 0.8 and the pixel size to 0.06$''$. The resulting images cover an area of 3$'\times$3$'$, with image qualities\footnote{The image quality reported in Table 1 corresponds to the FWHM of the intrinsic PSF resulting from the deconvolution of the observed PSF and the pixel response function of the WFC3 IR detector \citep{Koekemoer_2011}.} that vary between 0.11$''$ and 0.14$''$. The WFC3 observations of XMM1229 are summarised in Table 1.

\subsection{Ground Based Imaging}

\subsubsection{FORS2}

XMM1229 was observed with the FOcal Reducer/low dispersion Spectrograph 2
\citep[FORS2, ][]{Appenzeller_1998}, mounted on {\itshape{Yepun}}, the fourth
unit of the 8 m ESO/VLT, on Cerro Paranal (Chile). The observations were taken during
program 073.A-0737(A) (P.I. A. Schwope), an optical follow-up of the XDCP,
carried out with the {\ttfamily R\_SPECIAL} and {\ttfamily Z\_GUNN} filters. We only used the {\ttfamily R\_SPECIAL} (R)
data, as the quality of the ACS $z_{850}$ image is significantly higher. The FORS2 camera
comprises a detector array of two 2k $\times$ 4k MIT CCDs, separated by a 7
pixel gap and, in order to correct for that, the XMM1229 field was observed in
three dithered exposures of 380 s each, that were reduced using the FORS data
reduction pipeline{\footnote{http://www.eso.org/sci/software/pipelines/.}}.
Each chip was separately bias subtracted and flat-field corrected, and a standard
star field was used to estimate the zero point. The Software for Calibrating AstroMetry and Photometry (SCAMP, \citealt{Bertin_2006}) was then used to find an appropriate
astrometric solution for the images, that were finally co-added using SWarp
v2.19.1 \citep{Bertin_2002}{\footnote{The SCAMP and SWarp binary and source files can be found at http://www.astromatic.net.}.
The field of view of the final R band image of XMM1229 is 8$'$ $\times$ 9$'$ with an
image quality of $\sim$0.6$''$ and a resolution of 0.252$''$/pixel. The
FORS2 observations are summarised in Table 2.

\subsubsection{SofI}

XMM1229 was observed in the J and Ks filter bands of SofI \citep{Moorwood_1998}, mounted on the ESO New
Technology Telescope (NTT), at the La Silla Observatory (Chile). The data were taken in March 2007 as part of a near
infrared (NIR) follow-up of the XDCP. In this paper we only use the J band, as
the Ks data from HAWK-I are considerably deeper (see \S 2.2.3 and Table 2). We summarise
here the SofI observations of the XMM1229 field, referring to \cite{Santos_2009}
for a more detailed description. 

XMM1229 was observed for a total of 40 minutes
with the instrument operating in {\itshape Large Field Mode}, which has a 5 $'$
$\times$ 5 $'$ field of view, with a resolution of 0.290$''$/pixel. In order to
take into account the high variability of the NIR background, dithered exposures
of the field were taken and reduced with the ESO/MVM software
\footnote{
http://archive.eso.org/cms/eso-data/data-packages/eso-mvm-software-package}. The
resulting image quality is $\sim$0.9$''$. The SofI J band
observations are summarised in Table 2.

\subsubsection{HAWK-I}

A subsample of the HST Cluster Supernova Survey, one cluster from SpARCS \citep{Muzzin_2012, Lidman_2012} and the cluster RX J0152.7-1357, from the ACS Intermediate Redshift Cluster Survey \citep{Ford_2004}, all observable from the southern
hemisphere, were observed with the High Acuity Wide field K-band
Imager (HAWK-I, \citealt{Pirard_2004}), mounted at the Nasmith A focus of {\itshape{Yepun}}, the fourth
unit of the 8 m ESO/VLT. These clusters are part of the HAWK-I Cluster Survey (HCS) \citep{Lidman_2013}, a near-infrared program providing deep imaging data necessary for the study of the
properties of old stellar populations in cluster early-type galaxies at high redshift. We refer to \cite{Lidman_2013} for a 
detailed description of the HAWK-I observations and data reduction of the XMM1229 field. 

The HAWK-I camera consists of an array of 4 detectors, each defining the quadrant of a square surface, with a total area corresponding to a  7\farcm5 $\times$ 7\farcm5 field of view.  The final co-added mosaic of XMM1229 has a field of view of 10$'$ $\times$ 10$'$ with an image quality of 0.34$''$ and a resolution of 0.1$''$/pixel. The HAWK-I observations are summarised in Table 2.

\subsection{Ground-based Spectroscopy}

XMM1229 was observed with FORS2 in 2006 during the spectroscopic follow-up of SNe Ia in the HST Cluster Supernova Survey. We summarise here the observations and data reduction, referring the reader to \cite{Santos_2009} and \cite{Suzuki_2012} for more detailed descriptions. 

XMM1229 was observed five times with FORS2 using the 300I grism and the OG590 order sorting filter. This configuration produces a spectral resolution of 2.3 \AA/pixel and a wavelength coverage extending from 5900 to 10,000 \AA. Several SNe were detected in this cluster, and for this reason, and to allow the supernovae to be observed both near their maximum light and when their luminosity had significantly faded, XMM1229 was targeted multiple times. A total of 77 objects were observed, 26 of which were found to lie within 3$\sigma$ from the average cluster redshift ($z \sim 0.98$). The spectroscopically confirmed members of XMM1229 are represented as red dots in Fig. 4 and are listed in Table 3. Six galaxies were only detected in the HAWK-I image and we exclude them from the analysis of the red sequence, as no colour information is available. The galaxy XMM1229\_316 was also not considered, as SExtractor \citep{Bertin_1996} failed to return a reliable estimate of the $i_{775}$ and $z_{850}$ aperture 
magnitudes (SExtractor {\ttfamily{FLAGS}} = 16: data within the aperture incomplete or corrupted, see also Table 3).

\begin{table*}
  \caption{Summary of the HST observations of XMM1229.}
  \begin{minipage}{14 cm}
  \begin{tabular}{|c|c|c|c|c|c|c|}
  \hline
         & F775W ($i_{775}$) & F850LP ($z_{850}$) & F105W (Y) & F110W & F125W
(J) & F160W (H) \\
  \hline
  \hline
 \textcolor{black}{telescope/instrument} & \textcolor{black}{HST/ACS} & \textcolor{black}{HST/ACS} & \textcolor{black}{HST/WFC3} & \textcolor{black}{HST/WFC3} & \textcolor{black}{HST/WFC3} & \textcolor{black}{HST/WFC3} \\
 exposure time (s) & 4160 & 10940 & 1312 & 1112 & 1212 & 1112 \\
 imaged field & 5.1$'$ $\times$ 5.1$'$ & 5.1$'$ $\times$ 5.1$'$ & 3$'$ $\times$
3$'$ &  3$'$ $\times$ 3$'$ &  3$'$ $\times$ 3$'$ &  3$'$ $\times$ 3$'$ \\
 pixel scale \footnote{Drizzled pixel scale, see \S 2.1.1 and \S 2.1.2} ($''$/pixel) & 0.05 & 0.05 & 0.06 & 0.06 & 0.06 & 0.06 \\
 image quality (FWHM) \footnote{FWHM of the PSF modelled by PSFex, see \S 3.1} & 0.08$''$ & 0.09$''$ & 0.11$''$ & 0.11$''$ & 0.13$''$ &
0.14$''$ \\
\textcolor{black}{90\% completeness (mag)} & \textcolor{black}{25.0} & \textcolor{black}{25.0} & \textcolor{black}{23.0} & \textcolor{black}{23.2} & \textcolor{black}{23.3} & \textcolor{black}{23.5} \\
\hline
\end{tabular}
\end{minipage}
\end{table*}

\begin{table*}
  \caption{Summary of the ground based observations of XMM1229.}
  \begin{minipage}{8 cm}
    \centering
  \begin{tabular}{|c|c|c|c|}
  \hline
         & R\_SPECIAL (R) & J & Ks\\
  \hline
  \hline
 \textcolor{black}{telescope/instrument} & \textcolor{black}{VLT/FORS2} & \textcolor{black}{NTT/SofI}  & \textcolor{black}{VLT/HAWK-I} \\
 exposure time (s) & 1140 & 2280 & 11310 \\
 imaged field & 8$'$ $\times$ 9$'$ & 5$'$ $\times$ 5$'$ & 10$'$ $\times$ 10$'$\\ 
 pixel scale ($''$/pixel) & 0.252 & 0.290 & 0.10 \\
 image quality (FWHM) \footnote{FWHM of the PSF modelled by PSFex, see \S 3.1} & 0.63$''$ & 0.94$''$ & 0.34$''$ \\
 \textcolor{black}{90\% completeness (mag)} & \textcolor{black}{25.3} & \textcolor{black}{22.4} & \textcolor{black}{24.6} \\
\hline
\end{tabular}
\end{minipage}
\end{table*}

\section{Data Analysis and Measurements}

\subsection{Object Detection and PSF modelling}

We group the images according to the observing program and the instrument with
which they were observed. This subdivision produces five groups: ACS, WFC3,
FORS2, HAWK-I and SofI. We use a modified version of the GALAPAGOS IDL pipeline\footnote{http://astro-staff.uibk.ac.at/~m.barden/galapagos/}
\citep{Haeussler_2007} to run SExtractor in high dynamic range mode (HDR) on each set of observations and
optimise the number of detections. In fact, when run with low detection
thresholds, small detection areas and aggressive deblending,
SExtractor can improperly split a bright object into a number of sub-components. On the other hand, when run with high detection
thresholds and large detection areas, the software might fail in deblending two
closely separated objects that are therefore classified as a single source. The HDR
technique consists of two distinct runs of SExtractor: one to detect only the
brightest sources ({\ttfamily COLD} run) and the other to detect the faintest sources
and deblend objects with 
very close neighbours ({\ttfamily HOT} run). This is achieved by varying the parameters
{\ttfamily DETECT\_MINAREA}, {\ttfamily DETECT\_THRESH} and {\ttfamily DEBLEND\_MINCONT}, and setting smaller
detection areas, detection thresholds and deblending contrasts in the HOT
run, when faint sources and close neighbouring objects are detected. This procedure
produces two separate catalogues: one for the bright and the other for the faint
sources, that are eventually merged into a single catalogue. GALAPAGOS
implements an algorithm in which the {\ttfamily HOT} sources that are within a
certain distance from each {\ttfamily COLD} source are rejected. In fact, these sources are likely to be
the result of improper deblending  (e.g.\  substructures in bright nearby galaxies or star formation clumps in distant late-type galaxies) and would contaminate the final catalogue as spurious detections.

We used the PSF Extractor (PSFex) software (version 3.9), written by the
Terapix group \citep{Bertin_2011} to model the PSF in each image\footnote{PSFex can be downloaded at:\\ http://www.astromatic.net/software/psfex}. PSFex  selects
point sources in the half light radius vs flux plane using the SExtractor
parameters {\ttfamily FLUX\_RADIUS} and {\ttfamily FLUX\_APER} as diagnostics of those two quantities. The PSF is  modelled as a linear
combination of basis vectors that can be chosen either using each pixel as a
free parameter (pixel basis), as done in this work, or using actual point source
images (Gauss-Laguerre and Karhunen-Lo\`{e}ve bases). The FWHM of the PSFs for each photometric band are reported in the fifth row of Tables 1 and 2.

\begin{figure*}
  \includegraphics[width=17 cm]{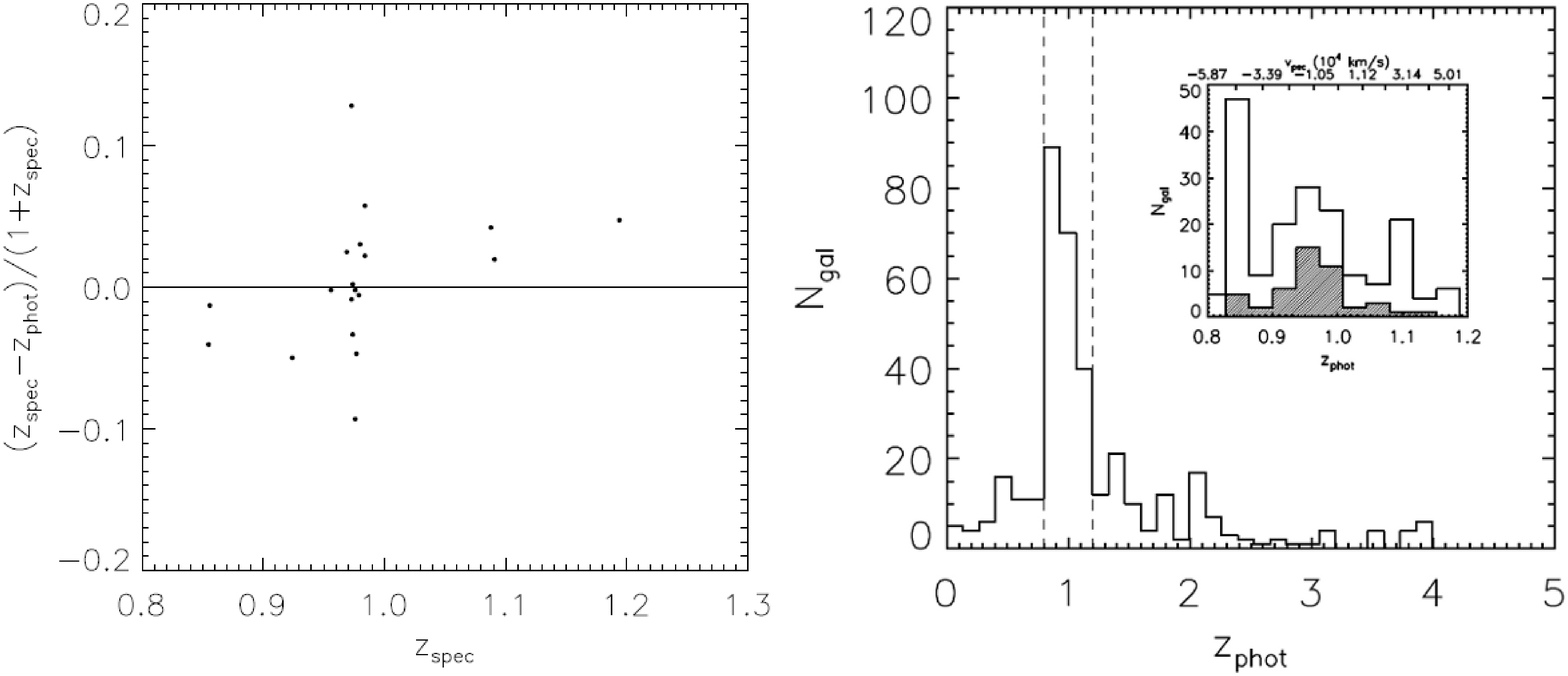}
	\caption{{\itshape{(left)}}: Calibration of the {\ttfamily{zpeg}} photometric redshift estimate for the cluster centre. On the x-axis it is plotted the value $z_{spec}$ of the spectroscopic redshifts measured with FORS2 in the cluster centre, while on the y-axis it is plotted the discrepancy $\Delta z = (z_{spec}-z_{phot})/(1+z_{spec})$. {\itshape{(right)}}: Photometric redshift distribution in the central region of XMM1229. The vertical dashed lines are the two limits used to define the cluster membership. The redshift distribution of the cluster members and of the cluster red sequence members (hatched histogram) is shown in the inset plot, where we also report the corresponding values of the galaxy peculiar velocities along the top horizontal axis.}
\end{figure*}

\subsection{Photometric Catalogue}

In order to study the properties of the members of XMM1229 across the whole
spatial extension of the cluster and given the different areas covered in each band pass, we decided to split our imaging database
into two samples: the centre and the outskirts of the
cluster. We produced a multiband catalogue for each sample. 

In this process we defined the central region as the one delimited by the WFC3 observed field, which
has the smallest extent. This provided us with 8 photometric bands in the
cluster centre (R, $i_{775}$, $z_{850}$, F105W, F110W, F125W, F160W, Ks) and 5 in
the cluster outskirts (R, $i_{775}$, $z_{850}$, J, Ks). We used the SofI J band
only for the analysis of the cluster outskirts because its spectral coverage
overlaps with the one of the WFC3 F125W which, being deeper, was used as the J
band of choice in the study of the cluster centre. The central region extends up to
600 kpc at the redshift of the cluster, which corresponds to $0.54 \times R_{200}$ \citep{Jee_2011}, while the outskirts region approximately corresponds to the region between 0.6 Mpc and 1.04 Mpc, the upper limit being imposed by the width of the ACS field.

The images of the cluster centre were degraded to the PSF of the R-band image, while in the outskirts they were degraded to the PSF of the SofI J band, which had the broadest FWHM. We  then ran GALAPAGOS on each of the PSF
matched images, using the unconvolved images for detection. With the HST
observations having been taken in more than one filter band, we used the $z_{850}$ and
F110W images for detection in the ACS and WFC3 groups, respectively. In order to remove any bias induced by intrinsic colour gradients, we measured
aperture magnitudes within fixed circular apertures of 2$''$, corresponding to
a physical radius $R_{ap} \sim 8$kpc at $z=0.98$. With this choice the colour gradients for bright galaxies become negligible, while at low luminosities galaxies are almost entirely contained within the aperture radius. Magnitudes were corrected for galactic extinction using the dust maps of \cite{Schlegel_1998} and applying the \cite{Cardelli_1989} equations.

In order to consistently compare galaxy colours in the cluster centre and outskirts\textcolor{black}{, following \cite{Meyers_2012}}, we
performed a cross-convolution of the (unconvolved) F775W and F850LP images, in which each
image was convolved by the PSF of the other image. These are in fact the two
photometric bands that were used for colour measurements  in this work, as they almost 
bracket the 4000 \AA\ break at the redshift of XMM1229 and are also the
deepest images of the sample (see Fig. 2). \textcolor{black}{The choice of this strategy for colour measurement allowed us to match the image qualities of the ACS images without degrading them to the ground-based level. In fact, after cross-convolution, the resulting image quality is $\sim 0.14''$, which is considerably narrower than in the ground-based images (see Table 2).}

The resampling and co-addition steps of the data reduction, as well as the PSF
matching process introduce correlations between pixels, which are not taken into
account by SExtractor \citep[see e.g.][]{Casertano_2000, Lidman_2008, Trenti_2011}. Following the method outlined in \cite{Labbe_2003}, we took random
regions of sky on each image and measured the flux within different apertures. This provided us with a direct measure of the variation of the sky flux with the aperture radius and therefore allowed us to quantify the deviations from a purely poissonian noise. The advantage of this method, with respect to the application of analytical relations, as those outlined in \cite{Casertano_2000}, is that the noise is estimated directly on the images and no assumptions are made about the properties of the instrument or the co-addition and resampling algorithms used in data reduction.
 
\textcolor{black}{In order to quantify the depth of the images, we inserted simulated galaxy images, generated as described in \S 4.4.3, in empty regions of each science image. We ran SExtractor on each single image in single image mode with the same configuration used for the original images and looked at the fractions of recovered simulated objects as a function of input magnitude. This fraction is a direct measurement of the incompleteness of the photometric catalogues extracted on each image. The 90\% magnitude completeness limit, quoted in the sixth row of Tables 1 and 2, is used in this work to parametrise completeness in the XMM1229 sample.}

\section{Results}

\subsection{Cluster Membership: Photometric Redshifts}

With the available FORS2 spectra, it is possible to study the red sequence only down to $z_{850}=23.0$ (see Fig. 4).
In order to estimate the distance of fainter galaxies and assess their membership to XMM1229, we could rely either on statistical background subtraction or
photometric redshifts. These two methods are shown to produce comparable luminosity functions for red sequence cluster members \citep{Rudnick_2009}. With nine available photometric bands in the XMM1229 field, spanning the range $0.65<\lambda<2.2$ $\mu m$, corresponding to the rest-frame wavelength range $0.33<\lambda<1.1$ $\mu m$, covering from the near-ultraviolet to the near-infrared regions of the spectrum, and given the availability of 77 spectra, we decided to use photometric redshifts to determine the membership of the cluster. We were able to determine photo-z's for objects as faint as $z_{850}=24.0$ on the red sequence, going one magnitude fainter than the limit imposed by the FORS2 spectroscopic observations (see also \citealt{Santos_2009}) and remaining within the magnitude limit for a reliable morphological classification at $z\sim1$ \citep{Postman_2005}. 

The program {\ttfamily zpeg} \citep{Leborgne_2002} was used to fit synthetic spectral energy distributions (SED), grouped
in seven galaxy types, and built with the P\'EGASE spectral evolution code
\citep{Fioc_1997} assuming a \cite{Kroupa_2001} initial mass function (IMF). The template types cover a wide range of spectral classes, going from passive to actively star-forming galaxies. {\ttfamily zpeg} implements a $\chi^2$ minimisation procedure, in
which the best fitting SED is the one which minimises the $\chi^2$ in a
three-dimensional parameter space of age, redshift and template type. The metallicity of the synthetic SEDs is assumed to evolve with time according to the star formation history of each template and with the stars forming at the metallicity of the interstellar medium. No dust extinction is assumed. 

We used the available spectroscopic redshifts to calibrate the photo-z's obtaining a median $\Delta z = (z_{spec}-z_{phot})/(1+z_{spec}) = 0.0 \pm 0.05$ in the cluster centre and $\Delta z = 0.02 \pm 0.11$ in the outskirts, where the uncertainty is computed as the normalised median absolute deviation (NMAD, see also Fig. 3, left panel). In both samples we fixed $z_{phot}=z_{spec}$ for the spectroscopically confirmed cluster members and let {\ttfamily zpeg} perform the SED fitting with only age and template type as free parameters.

Cluster member candidates were defined as those galaxies in the range $0.8 < z_{phot} < 1.2$, the cut being
chosen from the width
of the photo-z distribution (see Fig. 3). \textcolor{black}{Although the median fractional photo-z error for all galaxies in the range $0.8 < z_{phot} < 1.2$ is 12 \%, we decided to focus only on the red sequence, as the photo-z estimate of blue galaxies is expected to be significantly uncertain with the available spectral coverage. In fact, the R band, which is the bluest for the XMM1229 samples, does not cover the blue side of the 4000 \AA\ break at $z<0.6$ and foreground galaxies may be misclassified as blue cluster members.} A second quantity produced by {\ttfamily{zpeg}} is the stellar mass, which is defined as the mass locked into stars and is obtained with a median fractional uncertainty of
$\sim$ 24\% for red sequence cluster members in the central region\footnote{The reader can refer to \cite{Bernardi_2010} for conversions to stellar masses obtained with the most commonly used Chabrier and Salpeter IMFs.}. The analysis of the stellar masses of red sequence galaxies in the cluster outskirts will be presented in a forthcoming paper.

\textcolor{black}{Delaye et al. (2013, submitted) studied the stellar mass vs size relation in the HCS clusters. They estimated stellar masses using the {\ttfamily{lephare}} software \citep{Arnouts_1999, Ilbert_2006} on a set of synthetic SEDs from the \cite{Bruzual_2003} library with three different metallicities ($0.2Z_\odot$, $0.4Z_\odot$, $Z_\odot$), exponentially declining star formation histories and a \cite{Chabrier_2003} IMF. As a consistency check, we estimated the stellar masses of red sequence galaxies in XMM1229 using {\ttfamily{lephare}} on the same set of templates of Delaye et al (2013) and adopting their cosmology\footnote{Delaye et al. (2013) use a $\Lambda$CDM cosmology with $H_0 = 70$ km $\cdot$ s$^{-1}$ $\cdot$ Mpc$^{-1}$, $\Omega_M = 0.30$, and $\Omega_{\Lambda} = 0.70$}. As in Delaye et al. (2013), we also fixed the redshift of red sequence galaxies at $z=0.98$ and we found that the so obtained stellar masses were a factor of 1.2 smaller. This small difference can be attributed to both 
the different sets of photometric bands used in the two works\footnote{Delaye et al. (2013) use $i_{775}$, $z_{850}$, $J$ (from SofI) and Ks.} and the different apertures used for galaxy photometry, as Delaye et al. (2013) used MAG\_AUTO magnitudes instead of fixed aperture magnitudes. When comparing with our {\ttfamily{zpeg}} estimates, we found a median ratio of 1.35 between Delaye et al. (2013) and this work. Since this difference did not affect the conclusions of this paper and in order to be consistent with the cosmology chosen for this work, we kept our stellar mass estimates.}

\subsection{Contamination from Field Interlopers}

\textcolor{black}{In order to estimate the contamination of the XMM1229 red sequence, we used the HST/ACS F775W and F850LP images of the two fields of the Great Observatories Origins Deep Survey (GOODS, \citealt{Giavalisco_2004}, version 2.0). These images are very similar to those of XMM1229, the only remarkable difference being their resolution ($0.03 ''$/pixel for GOODS vs $0.05 ''$/pixel for XMM1229).}

\textcolor{black}{GOODS is a deep astronomical survey centred on two fields: the Hubble Deep Field North (GOODS North) and the Chandra Deep Field South (GOODS South). The project was aimed at collecting deep multiband photometry from various space- and ground-based facilities (e.g.\ HST, Spitzer, Chandra and XMM-Newton, VLT, KPNO, Subaru), in order to accurately study the properties of distant galaxies. Spectroscopic follow-up observations were also carried out at Keck and VLT \citep{Wirth_2004, Vanzella_2005, Vanzella_2006, Vanzella_2008, Popesso_2009, Balestra_2010}, resulting in an extensive ensemble of imaging and spectroscopic data covering $\sim 300$ square arcminutes}

\textcolor{black}{As for XMM1229, we followed the method outlined in \S 3 to process the GOODS ACS images and model the PSF. The 90\% magnitude completeness limits, estimated as described in \S 3.2, are $i_{775, lim} = 27.3$ mag and $z_{850, lim} = 26.7$ mag, about two magnitudes deeper than in the XMM1229 field. The $i_{775}-z_{850}$ colours were again measured on the cross-convolved images within $2''$ fixed apertures.}

\textcolor{black}{We adopted the method outlined in \cite{Pimbblet_2002} to estimate the fraction of galaxies contaminating the XMM1229 red sequence, modifying the equation in Appendix A of their paper to take into account the galaxies with assigned spectroscopic redshift in the XMM1229 field. This method assigns to each galaxy within a certain range of magnitude and colour a probability of belonging to the field. As a result, this probability quantifies the amount of contamination of the observed colour-magnitude diagram in the cluster field. We divided the colour-magnitude planes of the GOODS and XMM1229 fields into cells of equal width in colour and magnitude and in each cell we computed the probability for each galaxy of being in the field:
\begin{equation}
P_{field} = \frac{N_{field} \times A - N_{XMM1229, cont}}{N_{XMM1229} - N_{XMM1229, cont}}
\end{equation}
where $N_{field}$ is the number of galaxies in each cell of the GOODS colour-magnitude diagram, $N_{XMM1229}$ is the number of galaxies in each cell of the XMM1229 observed colour-magnitude diagram, $N_{XMM1229, cont}$ is the number of galaxies in each cell of the XMM1229 observed colour-magnitude diagram which are known not to be in the cluster from their spectroscopic redshift, and $A$ is the ratio between the areas of the XMM1229 field and the GOODS field. The main drawback of this method is that one can have $P_{filed} > 1$ or $P_{filed} < 0$. As suggested by \cite{Pimbblet_2002}, in these cases the width of the cell is adjusted until $0.0 < P_{field} < 1.0$.}

\textcolor{black}{For the purpose of this paper we only focused on the red sequence and we split it into two cells at $21.0 \leq z_{850} < 22.5$ and $22.5 \leq z_{850} < 24.0$ with the same colour width $0.7 \leq (i_{775}-z_{850}) < 1.1$. This choice was motivated by the fact that the cells sample the observed red sequence with sufficient resolution without falling into a regime of excessive low-number statistics (see Fig. 4). Furthermore, the magnitude limits are the same adopted for the estimation of the luminous to faint ratio in \S 5. The average $P_{field}$ for the observed XMM1229 red sequence is $P_{field}\sim 6$\% in the cluster centre and $P_{field}\sim 32$\% in the outskirts.} 

\textcolor{black}{With the definition of field contamination given in Equation (1), the number of cluster members $N_{cluster}$ in each colour-magnitude cell is defined as:
\begin{equation}
N_{cluster} = (1 - P_{field})*(N_{XMM1229} - N_{XMM1229, cont}).
\end{equation}}
\textcolor{black}{We use this equation to correct for outliers in the estimation of the luminous-to-faint ratio (see \S 5).}

\subsection{Colour-Magnitude Diagram and Red Sequence}

The colour-magnitude diagrams for the cluster centre and outskirts are presented in figure 4. In order to model the red sequence, we applied a robust linear fit, implementing the Tukey's bi-square weight function, and restricting the fit to the photometrically confirmed members in the colour range $0.75 < (i_{775} - z_{850}) < 1.5$ at $z_{850} < 24.0$. The model red sequence was defined as:
\begin{equation}
(i_{775}-z_{850})_{RS} =  a + b \times (z_{850}-21.0)
\end{equation}
where $b$ is the slope and $a$ represents the colour of a galaxy on the red sequence at $z_{850}=21.0$. We obtained $b = -0.044 \pm 0.017$ and $a=0.94 \pm 0.03$ for the cluster centre, where the uncertainties on $a$ and $b$ were estimated by generating 1000 bootstrap samples from the photometrically confirmed members used to fit the red sequence. Following \cite{Lidman_2004} and \cite{Mei_2009}, we estimated the intrinsic scatter $\sigma_c$ of the red sequence as the scatter that needed to be added to the colour error to have $\widetilde{\chi}^2=1.0$, where $\widetilde{\chi}^2$ is the reduced $\chi^2$. We found $\sigma_c=0.026 \pm 0.012$, where the uncertainty was again estimated by creating 1000 bootstrap samples from the sample of photometrically confirmed members used in the red sequence fit. We discuss the implications of these results in \S 5.4.

\textcolor{black}{As shown in Table 2, the SofI J band 90\% completeness limit is $J=22.4$. This results into a loss of red sequence objects at magnitudes $z_{850} > 22.5$ in the photo-z selected outskirts sample. Therefore, in order to study the cluster outskirts. we first modelled the observed red sequence and then used Equation (2) to statistically subtract field interloper galaxies. We obtained the following result: $a=0.88\pm0.05$, $b=-0.01\pm0.03$ and  $\sigma_c=0.052 \pm 0.015$. It can be seen that the red sequence is shallower than in the cluster centre and has a larger intrinsic scatter, although the slopes and the intrinsic scatters for the two samples are still consistent.}

\textcolor{black}{In the cluster centre the red sequence population was defined as those galaxies with $- 4\sigma_c < (i_{775}-z_{850}) - (i_{775}-z_{850})_{RS} < + 7\sigma_c$, where the limits were chosen after visually inspecting the colour-magnitude diagram, as the most suitable to bracket the red sequence. For the cluster outskirts we used the boundaries $- 3\sigma_c$ and $+ 4\sigma_c$.}

We defined the total galaxy magnitude as the extinction-corrected SExtractor {\ttfamily MAG\_AUTO}, although we are aware that this quantity underestimates the total flux, especially for elliptical and lenticular galaxies. \cite{Graham_2005} proposed general aperture corrections based on galaxy light profiles. However, in order to apply them to systems composed by a bulge and a disc, like S0 galaxies, one should first perform a bulge-disc decomposition which at $z\sim1$ becomes highly uncertain due to the high sky contamination at low fluxes. 

The photometric selection of the red sequence in the cluster centre produced a sample of 45 galaxies to which we added the bright spectroscopically confirmed member XMM1229\_145 ($z_{850}=21.9$), falling just below the red sequence. This brought the final number of objects analysed in the cluster centre to 46. The red sequence members of the central region are listed in Table 3. We did not assign cluster membership to the single galaxies in the outskirts, as a detailed study of the cluster outskirts in the HCS will be presented in a forthcoming paper. For this reason in Table 3 we only report the 3 spectroscopically confirmed red sequence members in this subsample. Our sample almost doubles the number of red sequence galaxies analysed in \cite{Santos_2009}, \textcolor{black}{ who limited themselves to the spectroscopically confirmed cluster members. This allows us to study galaxy properties along the red sequence down to $z_{850}=24.0$ mag, i.e. 1 mag fainter than in that work}.

\begin{figure*}
    \includegraphics[trim=0.0cm 0.0cm 0.0cm 0.0cm, clip, width=8 cm]{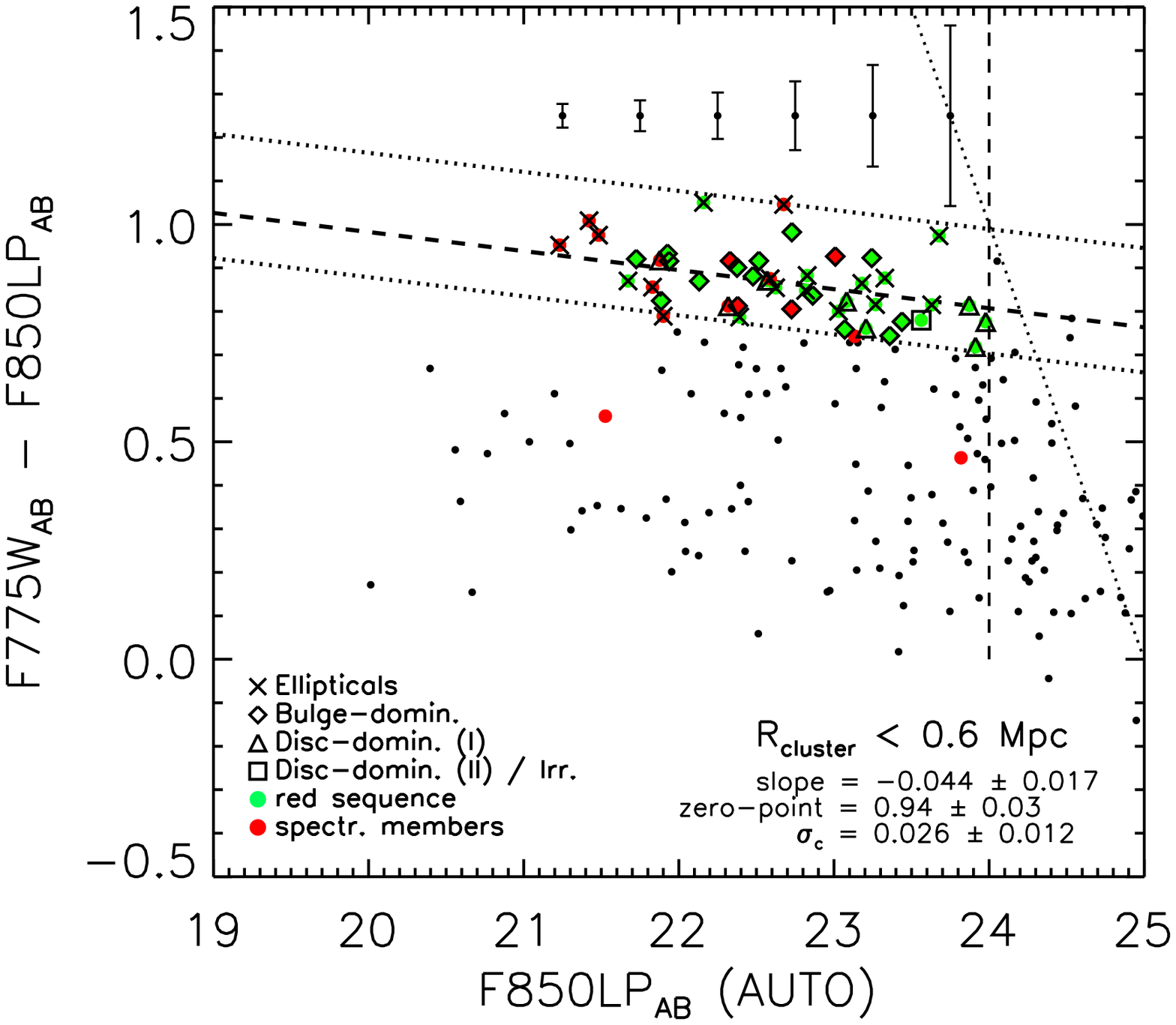}
    \hspace{5mm}
   \includegraphics[trim=0.0cm 0.0cm 0.0cm 0.0cm, clip, width=8 cm]{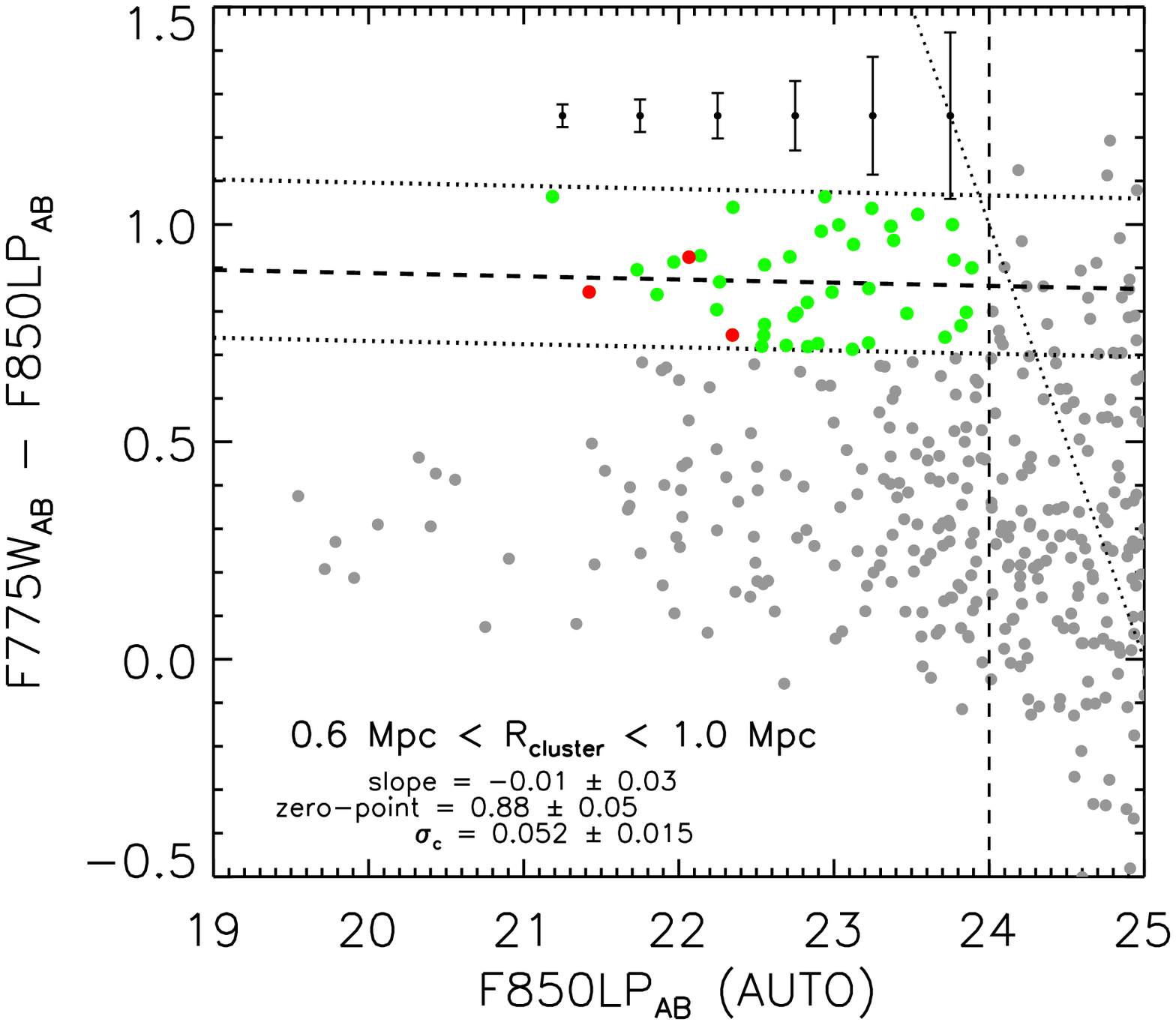}
	\caption{{\itshape (left)}: Colour-magnitude diagram of the central region of XMM1229. The colours are measured on the F775W and F850LP PSF cross-convolved images, adopting fixed circular apertures with 1$''$ radius ($\sim$8 kpc at $z=0.98$). Black dots represent all the photometrically selected cluster members and green dots are the red sequence members. Red dots are spectroscopically confirmed cluster members. The vertical dashed line represents the flux limit of visual morphology (see \S 4.4) and the sloping dotted line is the 90\% completeness limit in the F775W and F850LP images. The black dashed line is the linear fit to the red sequence and the two dotted lines represent the -4$\sigma_c$ and +7$\sigma_c$ envelopes delimiting the red sequence. The error bars represent the median colour errors along the red sequence in bins of 0.5 magnitudes. {\itshape (right)}: Observed colour-magnitude diagram in the outskirts of XMM1229. No cut in photometric redshifts is applied for this sample. Grey dots are all the galaxies observed in the cluster outskirts, red dots are spectroscopically confirmed cluster members and green dots are red sequence galaxies selected as described in \S 4.3. Error bars represent the median colour errors along the red sequence in bins of 0.5 magnitudes. The meaning of the lines is the same of the left panel.}
\end{figure*}

\begin{table*}
 \centering
  \caption{\textcolor{black}{Red sequence members of XMMU J1229+0151 (XMM1229). Red sequence members are photometrically selected as explained in \S 4.1 in the cluster {\itshape{centre}}. Spectroscopically confirmed members are indicated with their spectroscopic redshift and the estimated photometric redshift. The table is divided into the following sections: (1) red sequence members within 0.6 projected Mpc from the cluster centre ({\itshape{centre}} sub-sample); (2) spectroscopically confirmed red sequence members between 0.6 and 1.04 projected Mpc from the cluster centre ({\itshape{outskirts}}); (3) spectroscopically confirmed cluster members in the blue cloud or with corrupted photometry; (4) spectroscopically confirmed cluster members detected only in the HAWK-I Ks image. Objects in Sections (3) and (4) of this table are excluded from the analysis in this paper.}}
  \begin{tabular}{c|c|c|c|c|c}
  \hline
  \multicolumn{6}{|c|}{cluster centre}\\
  ID & $\alpha$ & $\delta$ & $z_{phot}$ & $z_{spec}$ & Morphology \\
  \hline
  \hline
      XMM1229\_128 & 12:29:28.315 &  +1:50:35.57 & 0.95 & & Disc-dominated (early) \\
      XMM1229\_145 & 12:29:29.944 &  +1:50:46.29 & 0.87 & 0.984 & Elliptical \\
      XMM1229\_172 & 12:29:27.718 &  +1:50:54.82 & 0.92 & 0.969 & Bulge-dominated \\
      XMM1229\_190 & 12:29:31.099 &  +1:51:1.23 & 1.07 & & Disc-dominated (early) \\
      XMM1229\_200 & 12:29:32.062 &  +1:51:5.37 & 0.94 & & Bulge-dominated \\
      XMM1229\_229 & 12:29:30.160 &  +1:51:16.37 & 0.92 & & Elliptical \\
      XMM1229\_237 & 12:29:29.297 &  +1:51:21.82 & 1.04 & 0.974 & Elliptical \\
      XMM1229\_240 & 12:29:29.420 &  +1:51:21.32 & 1.02 & & Bulge-dominated \\
      XMM1229\_241 & 12:29:24.819 &  +1:51:20.90 & 0.93 & & Elliptical \\
      XMM1229\_243 & 12:29:25.972 &  +1:51:20.26 & 0.84 & & Elliptical \\
      XMM1229\_244 & 12:29:29.028 &  +1:51:19.77 & 0.97 & & Elliptical \\
      XMM1229\_248 & 12:29:28.428 &  +1:51:21.29 & 0.99 & 0.979 & Bulge-dominated \\
      XMM1229\_255 & 12:29:31.043 &  +1:51:22.78 & 0.94 & 0.984 & Disc-dominated (early) \\
      XMM1229\_260 & 12:29:29.199 &  +1:51:25.77 & 0.98 & 0.976 & Elliptical \\
      XMM1229\_262 & 12:29:29.507 &  +1:51:24.20 & 1.11 & & Elliptical \\
      XMM1229\_263 & 12:29:26.924 &  +1:51:23.84 & 0.85 & & Disc-dominated (early) \\
      XMM1229\_265 & 12:29:28.929 &  +1:51:24.92 & 1.14 & & Elliptical \\
      XMM1229\_283 & 12:29:28.258 &  +1:51:30.16 & 0.90 & & Bulge-dominated \\
      XMM1229\_286 & 12:29:27.127 &  +1:51:28.46 & 0.83 & & Bulge-dominated \\
      XMM1229\_288 & 12:29:29.448 &  +1:51:28.81 & 1.07 & 0.977 & Bulge-dominated \\
      XMM1229\_291 & 12:29:33.193 &  +1:51:31.27 & 0.95 & & Bulge-dominated \\
      XMM1229\_306 & 12:29:28.803 &  +1:51:32.61 & 1.05 & & Disc-dominated (early) \\
      XMM1229\_309 & 12:29:29.665 &  +1:51:33.33 & 0.85 & & Bulge-dominated \\
      XMM1229\_310 & 12:29:27.661 &  +1:51:37.39 & 0.98 & & Bulge-dominated \\
      XMM1229\_312 & 12:29:28.714 &  +1:51:36.97 & 1.16 & 0.976 & Elliptical \\
      XMM1229\_319 & 12:29:30.026 &  +1:51:39.75 & 0.94 & & Bulge-dominated \\
      XMM1229\_320 & 12:29:32.176 &  +1:51:38.12 & 0.96 & & Elliptical \\
      XMM1229\_322 & 12:29:28.686 &  +1:51:39.12 & 0.98 & & Bulge-dominated \\
      XMM1229\_331 & 12:29:29.059 &  +1:51:40.51 & 0.89 & & Elliptical \\
      XMM1229\_353 & 12:29:27.637 &  +1:51:46.74 & 0.90 & & Elliptical \\
      XMM1229\_380 & 12:29:32.850 &  +1:51:52.51 & 0.93 & & Elliptical \\
      XMM1229\_392 & 12:29:29.704 &  +1:51:54.91 & 0.84 & & Disc-dominated (early) \\
      XMM1229\_394 & 12:29:27.774 &  +1:51:56.43 & 0.92 & & Elliptical \\
      XMM1229\_414 & 12:29:26.871 &  +1:52:0.98 & 1.07 & & Disc-dominated (late) \\
      XMM1229\_415 & 12:29:31.375 &  +1:52:3.62 & 0.97 & 0.969 & Elliptical \\
      XMM1229\_429 & 12:29:29.116 &  +1:52:4.92 & 0.87 & & Bulge-dominated \\
      XMM1229\_437 & 12:29:28.341 &  +1:52:6.73 & 0.96 & & Bulge-dominated \\
      XMM1229\_441 & 12:29:32.279 &  +1:52:6.90 & 0.99 & 0.973 & Bulge-dominated \\
      XMM1229\_456 & 12:29:27.302 &  +1:52:13.31 & 0.94 & & Disc-dominated (early) \\
      XMM1229\_457 & 12:29:28.330 &  +1:52:12.78 & 0.98 & & Bulge-dominated \\
      XMM1229\_463 & 12:29:32.598 &  +1:52:16.55 & 2.01 & 0.977 & Disc-dominated (early) \\
      XMM1229\_470 & 12:29:31.313 &  +1:52:15.97 & 0.96 & & Elliptical \\
      XMM1229\_475 & 12:29:29.264 &  +1:52:18.41 & 0.97 & 0.974 & Elliptical \\
      XMM1229\_477 & 12:29:28.867 &  +1:52:19.17 & 0.95 & & Bulge-dominated \\
      XMM1229\_502 & 12:29:30.523 &  +1:52:29.49 & 1.01 & & Bulge-dominated \\
      XMM1229\_287 & 12:29:29.187 &  +1:51:29.60 & 0.92 & 0.980 & Elliptical \\
   \hline
\end{tabular}
\end{table*}

\begin{table*}
 \centering
  \contcaption{}
   \begin{tabular}{c|c|c|c|c|c}
  \hline
  \multicolumn{6}{|c|}{cluster outskirts}\\
  ID & $\alpha$ & $\delta$ & $z_{phot}$ & $z_{spec}$ & Morphology \\
  \hline
  \hline
      XMM1229\_73 & 12:29:30.520 &  +1:50:11.02 & 0.95 & 0.979 & Elliptical \\
      XMM1229\_349 & 12:29:33.615 &  +1:51:46.39 & 1.02 & 0.973 & Bulge-dominated \\
      XMM1229\_373 & 12:29:33.260 &  +1:51:52.19 & 1.01 & 0.969 & Elliptical \\
      \hline
  \multicolumn{6}{|c|}{Spectroscopic members within ACS field excluded from present analysis}\\
  ID & $\alpha$ & $\delta$ & $z_{phot}$ & $z_{spec}$ & Morphology \\
  \hline
  \hline
      XMM1229\_183 & 12:29:23.202 &  +1:51:00.95 & 0.96 & 0.969 & Disc-dominated (early) (blue cloud) \\
      XMM1229\_308 & 12:29:32.951 &  +1:51:36.54 & 1.09 & 0.980 & Disc-dominated (early) (blue cloud)\\
      XMM1229\_487 & 12:29:29.643 &  +1:52:21.73 & 0.72 & 0.973 & Irregular (blue cloud) \\
      XMM1229\_316 & 12:29:25.778 &  +1:51:36.42 & ... & 0.968 & \textcolor{black}{Disc-dominated (late)} (SExtractor {\ttfamily{FLAGS}}=16, not used) \\
  \hline    
  \multicolumn{6}{|c|}{Spectroscopic members outside ACS field of view}\\
  ID & $\alpha$ & $\delta$ & $z_{phot}$ & $z_{spec}$ & Morphology \\
  \hline
  \hline
    FORS2\_4661 & 12:29:27.151 & +01:53:51.79 & ... & 0.975 & Elliptical \\
    FORS2\_4794 & 12:29:20.220 & +01:53:33.84 & ... & 0.974 & Bulge-dominated \\
    FORS2\_4800 & 12:29:16.481 & +01:53:41.62 & ... & 0.976 & Elliptical \\
    FORS2\_4910 & 12:29:17.113 & +01:53:33.19 & ... & 0.976 & Disc-dominated (early) \\
    FORS2\_4956 & 12:29:20.813 & +01:53:20.62 & ... & 0.978 & Elliptical \\
    FORS2\_5001 & 12:29:24.002 & +01:53:13.54 & ... & 0.973 & Elliptical \\
  \hline
  \end{tabular}
\end{table*}

\subsection{Galaxy Morphology and Structure}

We have shown in \S 4.2 that the outskirts of XMM1229 have a considerably large outlier contamination and therefore in this paper we restrict the morphological and structural analyses of red sequence galaxies to the cluster centre. The morphological and structural analyses of the cluster outskirts in the entire HCS sample will be the subject of a forthcoming paper.

\subsubsection{Morphological Classification}

We classified red sequence galaxies in the centre of XMM1229 using the ACS F850LP image, on which we were able to detect morphological features down to $z_{850}=24.0$. The F850LP filter corresponds approximately to a rest frame SDSS $g$ band, allowing us to compare directly with lower-redshift classifications performed either in the B or V bands (e.g.\ \citealt{Fasano_2012}). Galaxies were classified by three of the authors independently (P. C., W. J. C. and C. L.) on image cutouts whose size varied according to the SExtractor Kron radius of each object. We also ran the {\ttfamily galSVM} software (see \citealt{Huertas_2008, Huertas_2009a, Huertas_2011} and Appendix B) on the entire F850LP image, which provided us with a fourth independent and quantitatively-based classification. 

\textcolor{black}{Because at $z=1$  many higher-order morphological and structural features, such as spiral arms, bars and lenses are not resolved, we classified galaxies according to their observed bulge-to-total ratio (B/T) and split the red sequence sample into five broad morphological families: elliptical (E), bulge-dominated (BD), early disc-dominated (EDD), late disc-dominated (LDD) and irregular galaxies (Irr). A similar classification scheme was adopted by \cite{Postman_2005} and \cite{Mei_2012} to classify galaxies in ACS images of $z \sim 1$ clusters and it allows us to investigate the main structural features of cluster members.}

\textcolor{black}{Most S0s fall into the class of bulge-dominated galaxies, while the early and late disc classes comprise Hubble types going from Sa to Sbc and Sc to Scd, respectively. In the following we will use the terms `bulge-dominated' and S0 interchangeably, although we are aware that with our scheme the bulge-dominated sample may be contaminated by Sa galaxies (see e. g.\ \citealt{Mei_2012}) and the disc-dominated samples may be contaminated by low (B/T) S0 galaxies (S0$_c$ galaxies, \citealt{Laurikainen_2011}). In fact, spiral arms become fainter in the gas-poor galaxies which are typical of the red sequence population.}

\textcolor{black}{The four morphological classifications of the cluster centre agreed only for 4 galaxies, while 11 galaxies had the same type assigned in the three visual classifications. In particular, we note that while W. J. C. and C. L. agreed on 33 of the 46 red sequence galaxies, P. C. agreed with each author only on 16 and 17 galaxies, respectively. The main points of disagreement were the E/S0 distinction and the fact that faint elliptical galaxies tended to be classified by P. C. as disc-dominated systems with a very compact bulge and a faint disc. This underlines the challenges in morphological classifications of high redshift passive galaxies. For this reason we decided to adopt a majority rule for the assignment of the morphological types and thus the final morphological type was defined as the mode of the four independent classifications. There were four galaxies for which two of the classifiers assigned the type E and the other two the type BD, while in two other cases two of the classifiers 
assigned the type BD and the other two the type EDD. In these six cases we assigned type E to the 4 galaxies with no majority between type E and type BD and type BD to those with no majority between type BD and type EDD. For one galaxy (XMM1229\_322) two classifiers assigned type E and the other two type EDD. In this case we decided to classify the galaxy as bulge-dominated. Given the very low number of objects with late-type disc and irregular morphologies in both the XMM1229 and the two lower-redshift comparison samples (see \S 4.5), we decided to merge the two classes into one ``late disc-dominated / irregular'' morphological class. However, for completeness, in Tables 3, 4 and 5 we report the original morphological scheme with five classes.}

\textcolor{black}{The top panels of Fig. B1 show that, as expected, disc-dominated galaxies tend to have lower values of concentration and Gini coefficient with respect to elliptical and S0 galaxies (see e.g.\ \citealt{Lotz_2004}). We also note that the values of $M_{20}$ (lower left panel of Fig. B1) for disc-dominated galaxies are comparable with those of early-type galaxies. We interpret this result as a consequence of the fading of spiral arms in gas-poor spiral galaxies.}

The morphological classifications for the cluster centre and outskirts are reported in Table 3 (column 6), while thumbnail images for the galaxies listed in that table can be found in Fig. A1-A4. The morphology quoted for the cluster outskirts corresponds only to the output of {\ttfamily galSVM}.

\cite{Santos_2009} classified visually red sequence galaxies in XMM1229 using a scheme similar to ours: of the 15 galaxies in common with our cluster centre sample, there are 9 galaxies with the same assigned morphological type. When comparing only early- and late-type galaxies, we find that 13 of the 15 galaxies have the same assigned type. Delaye et al (2013) used {\ttfamily galSVM} to classify galaxies in the HCS clusters, dividing them into early- and late-type. The comparison with our classification shows that of the 45 galaxies in common to both samples, 38 (i.e. 84\%) were classified as early or late-type in both works. 

\begin{figure*}
 \vspace{5mm}
    \includegraphics[width=7.5 cm]{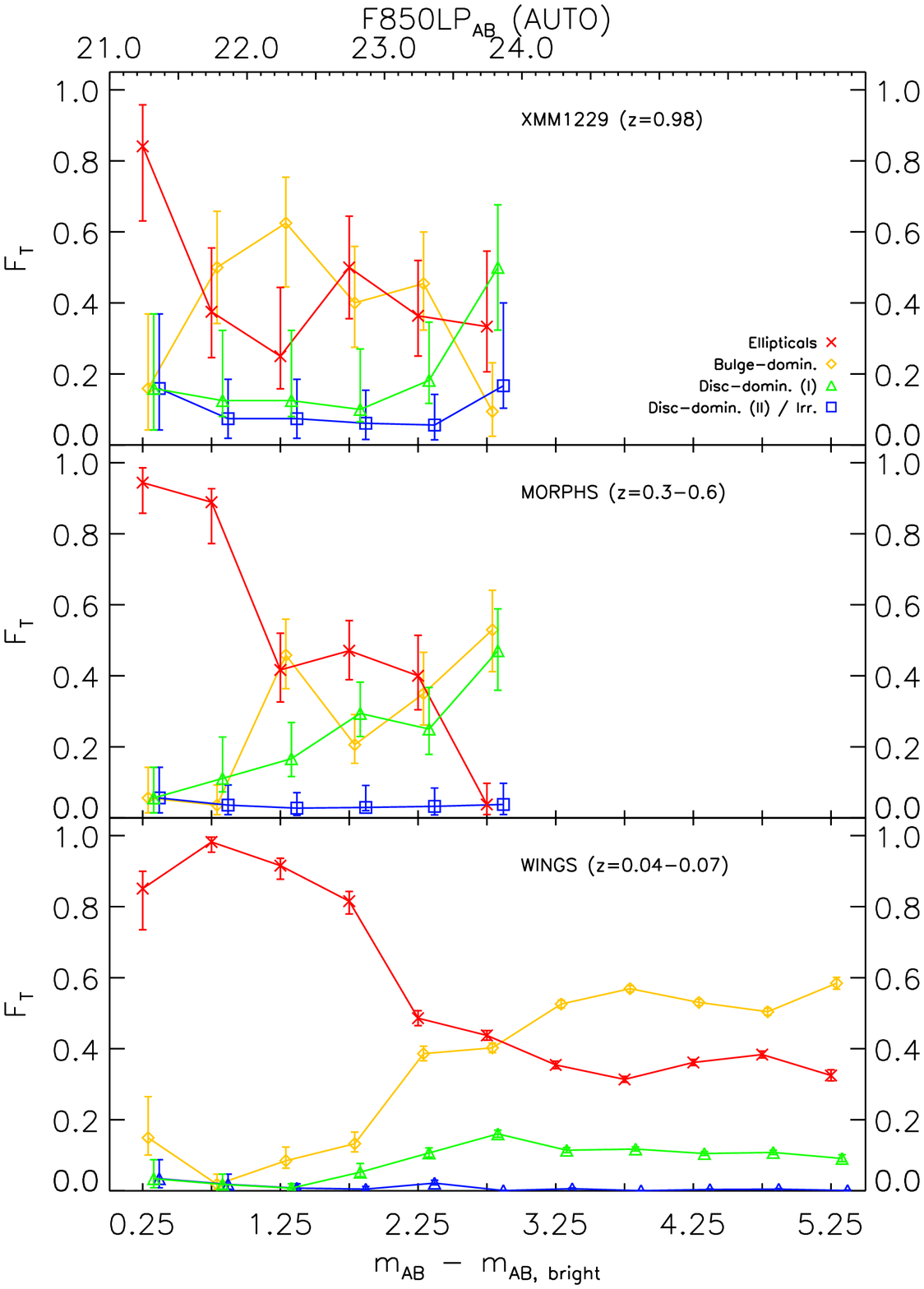}
    \hspace{5mm}
   \includegraphics[width=7.5 cm]{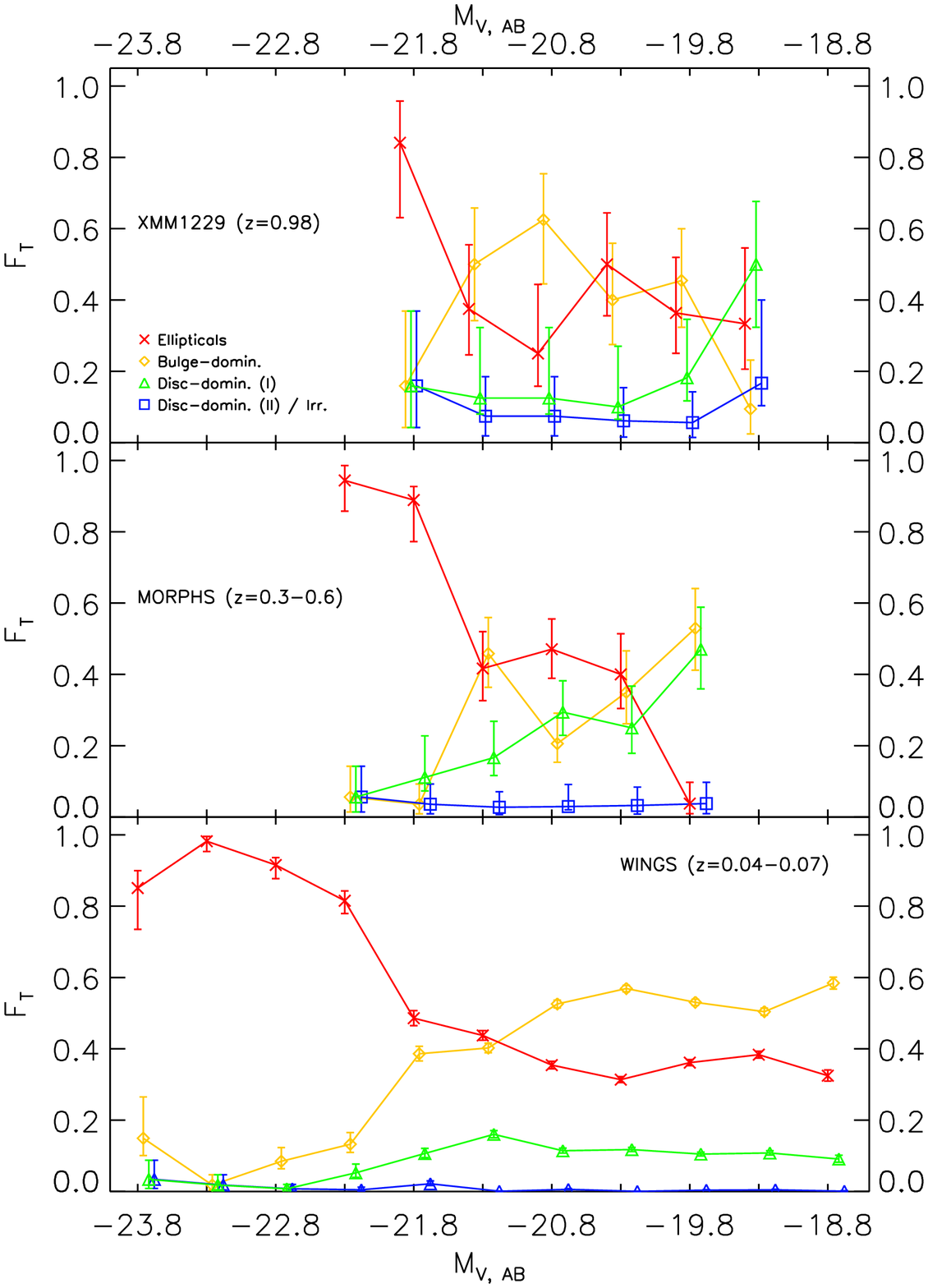}
	\caption{Morphological evolution of red sequence galaxies in clusters at $0.04 <  z < 0.98$. {\itshape{Left}}: magnitudes are normalised to the brightest bin in each sample. {\itshape (Top left panel)}: morphological fractions along the red sequence of XMM1229. {\itshape (middle left panel)}: fraction of morphological types along the red sequence of the spectroscopically confirmed MORPHS cluster members ($0.3 < z < 0.6$). {\itshape (bottom left panel)}: morphological fractions along the red sequence of the WINGS spectroscopically confirmed cluster members. The analysis in WINGS is restricted to objects with $V < 18.0$, corresponding to at least 50\% spectroscopic completeness. The bright end of the red sequence is dominated by elliptical galaxies as in MORPHS and XMM1229. At intermediate and low luminosities, S0 galaxies are the most frequent morphological class. On the top we report the apparent $z_{850}$ magnitude scale (down to the limit $z_{850}=24.0$ considered in this paper). {\itshape{Right}}: morphological fractions as a function of V absolute magnitude. {\itshape (Top right panel)}: XMM1229; {\itshape (middle right panel)}: MORPHS; {\itshape (bottom right panel)}: WINGS. In order to match the scales of the three samples, the magnitudes were passively evolved to $z=0$. The absolute magnitude axis is reproduced on the top of the plot. For clarity, in all the plots, the points for each morphological type are shifted along the x-axis by 0.04 mag.}
\end{figure*}

\subsubsection{The Morphological Composition of the Red Sequence.}

We divided the red sequence of the central region into bins of 0.5 magnitudes in the range
$21.0<z_{850}<24.0$ and, in each bin, we computed the fraction of the galaxy
population of a certain morphological type, $F_T$,  as:
\begin{equation}
F_{T,i} = \frac{N_{T,i}}{N_{tot,i}}
\end{equation}
where $F_{T,i}$, is the fraction of galaxies of type $T$ in the $i^{th}$ bin, $N_{T,i}$ is the number of galaxies of type $T$ in the $i^{th}$ bin, and $N_{tot,i}$ is the total number of galaxies in the $i^{th}$ bin. 

The error bars were computed following the method outlined in \cite{Cameron_2011} to estimate the confidence intervals for a binomial probability distribution. More precisely, in each magnitude bin, the confidence interval was estimated using a Bayesian approach in which the binomial probability mass function was treated as the {\itshape a posteriori} probability distribution, given a uniform prior over the expected number of successes. The errors on the morphological fractions were therefore estimated as the difference between the measured value of the fraction and the upper and lower bounds of the confidence interval. This method is shown to give reliable confidence intervals even with small samples, as in the case of this paper. Furthermore, it allows us to treat without ambiguity the extreme cases $F_T = 1$ and $F_T=0$. In fact, a gaussian approximation of the binomial distribution, suitable for large samples, would produce null errors, meaning certain estimates of the true value of the morphological 
fraction. We dealt with these two extreme cases defining as best estimate of $F_{T}$ the median of the a posteriori probability distribution and we estimated the errors as the difference between this value and the upper and lower bounds of the binomial confidence interval. 

The morphological fractions as a function of magnitude along the red sequence in XMM1229 are illustrated in the top panels of Fig. 5 and will be discussed in \S 5.

\subsubsection{Light Profile Fitting and Structural Parameters}

In order to investigate the connection between morphological and structural properties in red sequence galaxies, we fit a S\'ersic function to the light distribution of the red sequence members. We used GALFIT \citep{Peng_2002, Peng_2010_galfit}, implemented as part of the GALAPAGOS pipeline, on the whole F850LP image. The S\'ersic Law is parametrised by the equation:
\begin{equation}
\Sigma(r) = \Sigma_e e^{-\kappa \left[ (r/r_e)^{(1/n)} -1 \right]}
\end{equation}
where $\Sigma(r)$ is the galaxy surface brightness as a function of projected
radius, $r_e$ is the half light radius, $\Sigma_e$ is the surface brightness
at the half light radius, $n$ is the S\'ersic index and $\kappa$ is a
parameter which is coupled with $n$ in such a way that half of the total light
is always enclosed within $r_e$. The S\'ersic index is related to galaxy light
concentration: higher values of $n$ correspond to more concentrated light profiles.
Therefore, spheroidal galaxies are expected to have higher S\'ersic indices,
while disc-dominated galaxies are characterised by low values of $n$. When
$n=1$, eq.\ (4) assumes the form of the Exponential Law, which is typical of spiral galaxies, while when $n=4$, it assumes the form of the De Vaucouleurs Law \citep{DeVauc_1948}, typical of elliptical galaxies. 

\cite{Graham_2003} showed that the S\'ersic index correlates with the luminosity and stellar mass of elliptical galaxies, so that bright elliptical galaxies in  cluster cores may reach up to $n \sim 10$, while dwarf ellipticals may also have $n \sim 2$ (see also \ \citealt{Graham_1996}). Furthermore, \cite{Kormendy_1989} found that the De Vaucouleurs Law holds only for elliptical galaxies in a narrow magnitude range centred at $M_B=-21$. For these reasons, unlike \cite{Santos_2009}, who constrained the value of the S\'ersic index between 1 and 4, we kept $n$ as a free parameter when running GALFIT. 

In order to test the reliability of GALFIT, we inserted 580 simulated galaxy images in random spots
of the original F850LP image. The galaxies were built using the latest version of the
simulation script developed by the MEGAMORPH collaboration (kindly provided by Boris
H\"{a}u{\ss}ler; see also \citealt{Haeussler_2013}) and based on the method outlined in
\cite{Haeussler_2007}. The advantage of this method, with respect to other
more traditional software (e.g.\ IRAF {\ttfamily{mkobject}}), is that the inner regions of the
simulated galaxies are oversampled to take account of the higher curvature of
the light profile towards the centre. In order to fully investigate the performance of GALFIT, no correlation was assumed between $n$ and the luminosity of the mock galaxies. The results, after running GALAPAGOS with the same settings used for the original F850LP image, are shown in Fig. 6. It can be seen that, at $z_{850}$=24.0, GALFIT can still produce a reliable fit. From the bottom right panel of Fig. 6 it can also be seen that the output S\'ersic index appears slightly underestimated, although still consistent with the input value, at $n>3.5$. However, we note here that there are less 
simulated objects with these values of the S\'ersic index. The results of the S\'ersic fit for the red sequence members of XMM1229 are shown in Fig. 7 and are discussed in \S 5.3.

\begin{figure*}
	\includegraphics[width=14 cm]{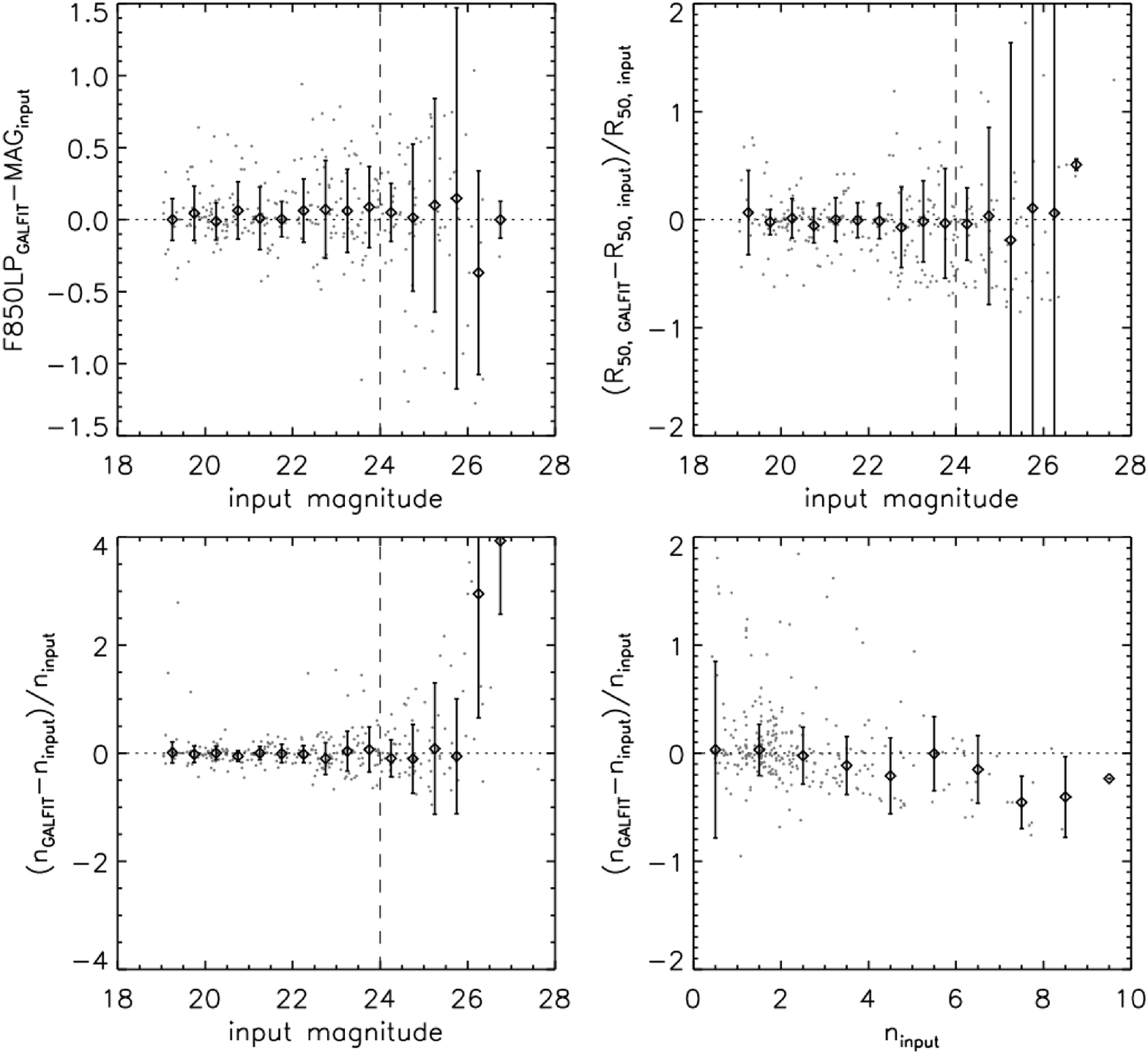}
	\caption{Performance of GALFIT on the F850LP image. Grey points represent the results for each retrieved simulated galaxy image, while black points and error bars are the median and half of the 68\% width of the distribution of each quantity in each bin on the x-axis. ({\itshape Top left}): galaxy magnitude: $\Delta z_{850}$ as a function of input magnitude; ({\itshape top right}): half light radius: $\Delta R_{50}$ as a function of input magnitude; ({\itshape bottom left}):  S\'ersic index: $\Delta n$ as a function of input magnitude; ({\itshape bottom right}): S\'ersic index: $\Delta n$ as a function of input S\'ersic index. The width of the input magnitude bins is 0.5 mag, while the width of the S\'ersic index bins is 1. The vertical dashed line represents the $z_{850} = 24.0$ limit for visual morphological classification (see \S 4.4.1).}
\end{figure*}

\subsection{The low-redshift Cluster Samples}

\subsubsection{MORPHS}

In order to compare with galaxy clusters at lower redshift, we built two comparison samples using the spectroscopic catalogues of the MORPHS and WINGS surveys. The MORPHS sample \citep{Smail_1997} comprises ten clusters in the redshift range $0.3<z<0.6$, observed with the Prime Focus Universal Extragalactic Instrument (PFUEI) on the 200 inch Palomar Hale telescope, and followed up with the Wide Field and Planetary Camera 2 (WFPC2) on HST. The clusters were observed in the $g$ and $r$ bands from the ground, while the WFPC2 images were taken in the F450W, F555W, F702W and F814W bands. Spectroscopic follow-up observations were conducted with the Carnegie Observatories Spectroscopic Multislit and Imaging Camera (COSMIC), on the Palomar 200-inch, the Low Dispersion Survey Spectrograph 2 (LDSS-2), on the William Herschel Telescope (WHT) at the Roque de los Muchachos Observatory (Spain), and the ESO Multi-Mode Instrument (EMMI) on NTT. The MORPHS spectroscopic observations and data reduction are described in \cite{
Dressler_1999}. The sample comprises in total 424 cluster members. 

The aim of the MORPHS project was to study the morphological and spectral 
properties of galaxies in clusters at intermediate redshifts; therefore an accurate visual morphological classification was conducted on the WFPC2 images of each cluster field \citep[see][]{Smail_1997}. The subsample of morphologically classified galaxies in the MORPHS spectroscopic sample comprises 122 cluster members. Since the redshift range of the MORPHS clusters allowed morphological features such as spiral arms or bars to be distinguished on the WFPC2 images, the classification scheme consisted of 9 types going from -7 to 10 and corresponding to the De Vaucouleurs T types \citep{DeVauc_1976}. In order to convert to the morphological scheme used for XMM1229, we grouped the galaxy morphological types and built broad classes similar to those used at $z\sim1$. The type conversion was performed following \cite{Poggianti_1999} and is summarised in Table 4. 

In order to identify red sequence galaxies in these clusters, we transformed all the photometry on to the rest-frame $B$ and $V$ system. In fact, only a part of the MORPHS WFPC2 observations were taken with a blue (F450W or F555W) and a red filter (F814W). The clusters observed with the F702W filter did not have any blue image and for them we had to resort to the ground-based $g$ and $r$ observations \citep{Dressler_1992}. We determined the k-correction using the software {\ttfamily zebra} \citep{Feldmann_2006} and a passive template SED from \cite{Coleman_1980}. To facilitate the comparison with XMM1229, we also converted the MORPHS magnitudes to AB magnitudes. We fitted the red sequence in the (B-V) vs V rest-frame colour-magnitude diagram using the same procedure adopted for XMM1229 and outlined in \S 4.3. We divided the red sequence into bins of 0.5 magnitudes each, in the range $-23.0 < V < -20.0$ and, in each bin, we measured the fractions of the single morphological 
types. The depth of the MORPHS spectroscopic observations does not allow us to investigate the faint end of the red sequence, however it constitutes a good sample to study the behaviour of the morphological fractions at bright luminosities. Since the spectroscopic follow-up of MORPHS was mainly focused on the study of the Butcher-Oemler effect, the target selection was biased towards bluer galaxies. Hence we corrected the morphological fractions using the normalisation factors given in Table 2 of \cite{Poggianti_1999}. The results are shown in the middle panels of Fig. 5 and are discussed in \S 5.2. Given the smaller population of the MORPHS spectroscopic sample, we did not restrict the morphological analysis of the red sequence to $R_{cluster} < 0.54 \times R_{200}$ as done for XMM1229 and WINGS (see \S 4.5.2).

\subsubsection{WINGS}

The WINGS cluster survey consists of 77 X-ray selected galaxy clusters in the range $0.04<z<0.07$, observed in the B and V bands with the Isaac Newton Telescope (Roque de los Muchachos Observatory) and MPG/ESO-2.2 m Telescope (La Silla). Spectroscopy from observations conducted at the WHT and the 3.9 m Anglo-Australian Telescope (AAT) is available for a subsample of 48 clusters with a total of 6000 redshifts \citep{Cava_2009}. Given the diverse spectroscopic coverage of each cluster and the consequent impossibility of studying the clusters individually, we built a composite sample taking all the spectroscopically confirmed members in each WINGS cluster with apparent magnitude $V<18.0$. In fact, according to Fig. 5 of \cite{Cava_2009}, the spectroscopic success rate in this flux region is greater than 50\%. We took into account the incompleteness 
of the sample along the red sequence weighing each galaxy by the inverse of the spectroscopic success rate at its magnitude. \textcolor{black}{A near-infrared (NIR) follow-up for 28 WINGS clusters was carried out by \cite{Valentinuzzi_2009} in the J and K band of the Wide Field Camera (WFCAM) at the UK Infrared Telescope (UKIRT) in Hawaii.}

In order to compare the same regions in units of $R_{200}$, we restricted the analysis of the WINGS clusters to $0.54 \times R_{200}$, corresponding to the same physical area covered by the XMM1229 centre sample.

After converting on to the AB system and deriving B and V rest-frame absolute magnitudes, we fitted the red sequence applying the same method used for  XMM1229 and MORPHS. The WINGS morphological classification was taken from \cite{Fasano_2012}, who used an approach based on neural networks. Their morphological scheme consists of 18 types that we grouped into broad classes to match the classification used for XMM1229 (see Table 5). The morphological composition of the red sequence in the WINGS cluster sample is shown in the bottom panels of Figure 5 and will be discussed in \S 5.2.

\begin{table}
 \centering
  \caption{MORPHS visual classification and type conversion (See \S 4.4.1 for details on the morphological scheme adopted for XMM1229).}
  \begin{tabular}{|c|c|c|}
  \hline
  T type & Morphology & XMM1229 morphology \\
  \hline
  \hline
 -7 & D/cD  &  Ellipticals \\
 -5 & E &  \\
  \hline
 -2 & S0 &  Bulge-dominated \\
  \hline
  1 & Sa &  Disc-dominated (early) \\
  3 & Sb &  \\
  5 & Sc &  \\
  \hline
  7 & Sd &  Disc-dominated (late) \\
  9 & Sm &         \\
  \hline
  10 & Irr & Irregulars \\
\hline
\end{tabular}
\end{table}

\begin{table}
 \centering
  \caption{WINGS morphological classification and type conversion (See \S 4.4.1 for details on the morphological scheme adopted for XMM1229).}
  \begin{tabular}{|c|c|c|}
  \hline
  type & Morphology & XMM1229 morphology \\
  \hline
  \hline
    -6 & cD & Ellipticals \\
    -5 & E & \\
    -4 & E/S0 \\
     \hline
    -3 & S0$^-$ &Bulge-dominated \\
    -2 & S0 & \\
    -1 & S0$^+$ & \\
     0 & S0/a & \\
    \hline
     1 & Sa & Disc-dominated (early) \\
     2 & Sab & \\
     3 & Sb & \\
     4 & Sbc & \\
     \hline
     5 & Sc & Disc-dominated (late) \\
     6 & Scd & \\
     7 & Sd & \\
     8 & Sdm & \\
     9 & Sm & \\
     \hline
    10 & Im & Irregulars \\
    11 & compact Im & \\
    
\hline
\end{tabular}
\end{table}

\section{Discussion}

\subsection{The Faint End of the Red Sequence}

Most authors quantify the galaxy deficit at the faint end of the red sequence with the ratio between the numbers of luminous and faint galaxies: $N_{lum}/N_{faint}$, also known as the luminous-to-faint ratio. This is equivalent to constructing the red sequence luminosity function dividing the sample into only two magnitude bins. The most used convention is to consider as luminous all the red sequence members with $M_V < -20.0$ and as faint all the red sequence members with $-20.0 \leq M_V < -18.2$, where the absolute magnitudes are measured in the Vega system and are passively evolved to $z=0$. This particular subdivision was first proposed by \cite{De_Lucia_2007} for galaxy clusters at $0.4 < z < 0.8$ and was motivated by the fact that their data reached $5\sigma$ completeness at $M_V = -18.2$, while $M_V = -20.0$ corresponded to about the midpoint of the red sequence. These limits allowed them to study the red sequence down to four magnitudes fainter than the BCG in each cluster, and could easily be 
applied to 
lower-redshift samples (see e. g. \citealt{Gilbank_2008, Capozzi_2010, Bildfell_2012}). 

In order to estimate $N_{lum}/N_{faint}$ in XMM1229, we had to convert our AB $z_{850}$ apparent magnitude to Vega V-band absolute magnitudes. \textcolor{black}{For this purpose we used a \cite{Bruzual_2003} simple stellar population (SSP) model with solar metallicity, formation redshift $z_f=4.75$ and exponentially declining star formation with e-folding time $\tau=1$ Gyr. We used the k-correction estimated for this model by the \cite{Bruzual_2003} colour and magnitude evolution software and converted to V-band magnitudes. The obtained V-band absolute magnitudes were finally passively evolved to $z=0$. We also tried exponentially declining star formation models with $z_f=3$ and $z_f=4$, as well as SSP models with an initial burst of star formation and $z_f=3,4,4.75$, but we found that the model with exponentially declining star formation and $z_f=4.75$ was the best in reproducing the average observed $i-z$ colour at $z=0.98$.}  After the conversion to Vega V-band magnitudes, the $21.0<z_{
850}<24.0$ range over which we study the properties of the red sequence in XMM1229 was mapped on to the $-22.0 < M_V <-19.0$ absolute magnitude range. We note that our V-band limit is about one magnitude brighter than that adopted by \cite{De_Lucia_2007} and other authors at lower redshifts. However, if we considered magnitudes fainter than $z_{850}=24.0$, we would fall into the incompleteness region of the colour-magnitude diagram (see Fig. 4). Thus, 
we fixed the boundary between bright and faint galaxies at $M_V=-20.5$ $(z_{850} = 22.5)$, which corresponds to the midpoint of the XMM1229 red sequence and measured $N_{lum}/N_{faint}$ over the $-22.0 < M_V <-19.0$ magnitude range. \textcolor{black}{With this definition, we obtain $N_{lum}/N_{faint} = 0.70 \pm 0.15$.}

\textcolor{black}{We measured the luminous-to-faint ratio on the composite spectroscopic WINGS sample applying the same V-band absolute magnitude cuts and obtaining:  $N_{lum}/N_{faint} = 0.364 \pm 0.006$. The errors on $N_{lum}/N_{faint} $ were estimated again following \cite{Cameron_2011}. Although $N_{lum}/N_{faint}$ is higher in XMM1229, as described below, the expected cluster-to-cluster scatter in WINGS is large, so we cannot conclude that there is evidence of evolution in $N_{lum}/N_{faint}$ based on this initial sample. If we use statistical background subtraction to correct for contamination instead of photometric redshifts, we obtain $N_{lum}/N_{faint} =0.68 \pm 0.13$, which does not change our conclusions.}

The existence of a deficit at the faint end of the red sequence supports the notion that less massive galaxies settle on to the red sequence at later epochs. As found by \cite{Demarco_2010} at $z=0.84$, the faint end of the cluster red sequence is populated by young low-mass early-type galaxies that had recently ceased to form stars. 

However, this scenario is still questioned by the studies of \cite{Andreon_2008} and \cite{Crawford_2009}, who found little or no decrease in the number of faint galaxies in clusters up to $z=1.3$. These authors attributed the observed deficit to measurement errors or cluster-to-cluster variations, claiming that the deficit may be a phenomenon that affects only some clusters. From Fig. 4 in \cite{Andreon_2008}, we find that at $z<1$, there is mild correlation between $\log{(N_{lum}/N_{faint})}$ and $\log{(1+z)}$ (Spearman coefficient $\rho=0.45$). At $z>1.0$, $N_{lum}/N_{faint}$ appears constant. We note that for these clusters the red sequence was selected using the ACS  $i_{
775}$ and $z_{850}$ bands, which do not bracket the 4000 \AA\ break at those redshifts. Therefore, the measurements of \cite{Andreon_2008} may be affected by fore- and background contamination of the red sequence sample. On the other hand, \cite{Capozzi_2010}, collecting different samples of clusters at $z<0.8$, found a significantly stronger correlation ($\rho=0.89$). 

\textcolor{black}{Cluster-to cluster variations play an important role in the study of the build-up of the red sequence and they need to be taken into account especially when dealing with small samples. \cite{Valentinuzzi_2011} estimated $N_{lum}/N_{faint}$ for 72 clusters in the WINGS photometric sample individually, and using statistical background subtraction to correct for interlopers. They used the same magnitude limits defined in \cite{De_Lucia_2007} and from the analysis of the distribution of their results, presented in their Fig. 4, we find ${(N_{lum}/N_{faint})}_{median}=0.5 \pm 0.2$, where the uncertainty corresponds to one half of the 68\% width of the $N_{lum}/N_{faint}$ distribution. This places XMM1229 in the upper tail of the $N_{lum}/N_{faint}$ distribution of WINGS.} 

\textcolor{black}{Since the measurements of $N_{lum}/N_{faint}$ for individual clusters are characterised by large uncertainties, particularly at high redshift (see also \citealt{De_Lucia_2007}), comparisons using a single cluster need to be treated with caution. In order to use the luminous-to-faint ratio to detect any build-up of the red sequence at low masses in galaxy clusters and to measure any trends with redshift, as done by \cite{Capozzi_2010}, \cite{Bildfell_2012} and \cite{Andreon_2008}, one needs to study a statistically significant number of clusters. In a forthcoming paper, we will use the complete HCS sample to study the existence of any correlation of $N_{lum}/N_{faint}$ with redshift extending in this way to higher redshifts the works of \cite{Capozzi_2010} and \cite{Bildfell_2012}, and looking at those epochs where \cite{Andreon_2008} found no correlation.}

\textcolor{black}{We also estimated $N_{lum}/N_{faint}$ in the cluster outskirts finding: $N_{lum}/N_{faint} = 0.35 \pm 0.11$, which is consistent with no deficit of faint galaxies at large distances from the cluster centre. However, we note that this result is still consistent to within $2\sigma$ with $N_{lum}/N_{faint}$ in the cluster centre.}

\begin{figure*}
  \includegraphics[width=17 cm]{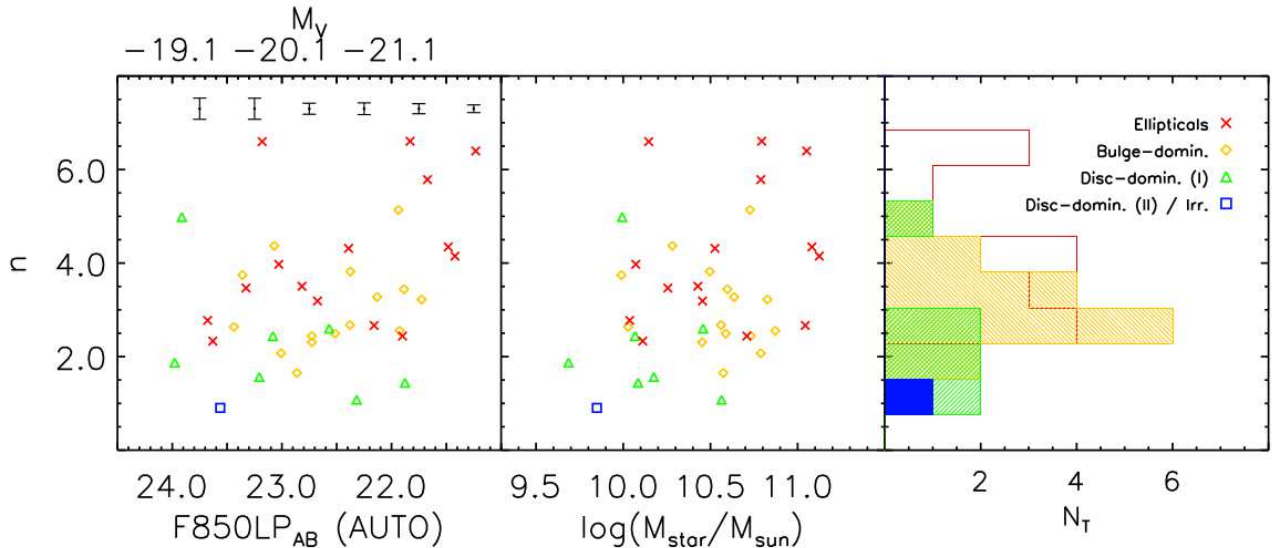}
     \caption{{\itshape (Left panel)}: S\'ersic index versus $z_{850}$ magnitude along the red sequence. Galaxies with later morphological types tend to have lower S\'ersic indices. The error bars represent the median GALFIT errors on the S\'ersic index in bins of 0.5 magnitudes. On the top we also report the Vega $M_V$ passively evolved magnitudes (see \S 5.2 for details). {\itshape (Central panel)}: S\'ersic index vs stellar mass along the red sequence. There is weak correlation between S\'ersic index and stellar mass and disc-dominated galaxies tend to have lower masses. {\itshape (Right panel)}: distribution of the values of the S\'ersic index for red sequence galaxies.}
\end{figure*}

\subsection{Morphological Evolution}

The left panel of Fig. 5 illustrates the comparison between the trends of the morphological fractions along the red sequence in XMM1229 and in the MORPHS and WINGS composite samples. In order to match the magnitude scales of the different samples, we offset the magnitudes to the brightest bin in each sample. In this way, we suppressed second-order effects such as the mismatches introduced by luminosity evolution and residual differences in photometry and k-correction, leaving only the magnitude differences between galaxies along the red sequence. This allowed us to compare more easily the trends of morphology along the red sequence, providing at the same time a clearer picture of the luminosity distribution for each morphological type. However, for completeness, in the right-hand panel of the figure, we also plot the morphological fractions as a function of absolute V-band magnitude passively evolved to $z=0$.

The top left panel of Fig. 5 shows that in XMM1229 the red sequence is mostly populated by elliptical and S0 galaxies. At the bright end, the red sequence is dominated by ellipticals, while going towards lower luminosities, the fraction of S0s increases and this class becomes predominant. At $22.5 < z_{850} < 23.5$ ($-20.9 < M_{V} < -19.9$) the fractions of ellipticals and S0s are comparable. Interestingly, we also see that {\textcolor{black}{at the same magnitudes}} there is a slight increase in the fraction of disc-dominated galaxies. At $z_{850} > 23.5$, the fraction of disc-dominated galaxies becomes dominant. 

The comparison with the WINGS sample shows that, at the relative magnitudes corresponding to the faint end in XMM1229, the red sequence is dominated by elliptical and S0 galaxies. We note a dearth of S0 galaxies in WINGS at the relative magnitudes in which S0s dominate the red sequence in XMM1229. However, it should be noted that this region of the red sequence is up to two magnitudes brighter in WINGS with respect to XMM1229 (see right-hand panel of Fig. 5). At similar absolute magnitudes the fractions of elliptical and S0 galaxies in the two samples look similar. This suggests that the bright end of the red sequence underwent significant evolution in the last 8 Gyr, becoming almost only populated by elliptical galaxies probably formed as a result of subsequent dry mergers (see \citealt{Faber_2007, Lidman_2013}). \cite{Vulcani_2011} studied the morphological fractions as a function of stellar mass in WINGS and in the clusters of the ESO Distant Cluster Survey (EDisCS, \citealt{White_2005}), at $0.4<z<0.8$. 
These 
authors excluded the BCGs from their analysis and still found a mild increase in the fraction of elliptical galaxies at low redshift, at stellar masses above $10^{10.9} M_\odot$.

\textcolor{black}{The large difference in V-band luminosity between WINGS and the two higher-redshift samples overpredicts the mass growth of the brightest cluster galaxies by a factor of 2 \citep{Lidman_2012}. In order to test the reliability of our estimate of the WINGS V-band absolute magnitudes, we looked at the $(V-K)$ vs $K$ colour-magnitude diagram for all the WINGS clusters with V and K band data. In fact, the K band is a good tracer of old stellar populations and therefore it is a good proxy of stellar mass, while the V band is more sensitive to young stars. The adopted $z_f = 4.75$ model with exponentially declining star formation history predicts $(V-K)_{Vega} \sim 3.3$ at $z=0.055$, the median redshift of the WINGS clusters. We found that at $V_{Vega}<-22.0$, for some of the clusters, the slope of the $(V-K)_{Vega}$ red sequence flattened or turned positive, making the average colour of red sequence galaxies bluer. A flattening of the WINGS red sequence at $V_{Vega} < -21.5$ in the $(B-V)$ vs $V$ 
colour-magnitude diagram was also mentioned by \cite{Valentinuzzi_2011}, although the authors did not discuss it. Interestingly, \cite{Bower_1992} studied the (V-K) colour of galaxies in the Virgo and Coma clusters finding $(V-K)_{Vega} \sim 3.3$ at 
the bright end of the red sequence, consistent with model predictions. The change in slope of the red sequence in these clusters and the resulting high V-band luminosities can be the consequence of residual colour gradients or systematics in the WINGS photometry such as PSF variations or differences in the processing of the B, V and K band data. However, some authors found that cooling flows can induce star formation in BCGs \citep{Crawford_1999, Rafferty_2008}, while simulations by \cite{Jimenez_2011} show that dry mergers between luminous red sequence galaxies and lower mass companions can result in more massive galaxies with lower total metallicity. These processes would all result in bluer colours at the bright end of the red sequence.}

\textcolor{black}{However, we point out that there are still big uncertainties in models of galaxy evolution. As outlined in detail by \cite{Skelton_2012}, the purely passive evolution of the red sequence, used in this work to compare clusters at different redshifts, underpredicts the brightness of the galaxies at low redshift, especially at the bright end of the red sequence. In particular, those authors point out that the concomitant effects of merger and star formation can explain both the slower evolution of the red sequence over the redshift range $0.0 < z < 1.0$ and the flattening of its bright end. Therefore, in order to rigorously compare red sequence galaxies in clusters at different redshifts, both merger and star formation must be taken into account and the passively-only evolved absolute magnitudes reported in the right-hand panel of Fig. 5 should be considered as upper limits to the $z=0$ analogues of the three cluster samples.}

The faint end of the red sequence in XMM1229 is characterised by the increase in the fraction of disc-dominated galaxies. At similar absolute magnitudes (i.e. $-20 < M_V < -19.0$), the WINGS red sequence is dominated by S0 and elliptical galaxies. This suggests that the disc-dominated galaxies observed in XMM1229 may be the progenitors of the S0 galaxies observed at similar absolute magnitudes in WINGS. Such a transformation could be the result of the stripping of gas from spiral galaxies that are falling into the cluster and are depleted of their gas reservoirs (strangulation). However, \cite{Bekki_2011} found that tidal interactions, due to slow encounters with other cluster members, can trigger bursts of star formation in the bulge of spiral galaxies, increasing the bulge fraction and transforming gas-rich spiral galaxies into gas-poor S0 galaxies. Owing to the recently formed stars, these galaxies should also have younger stellar ages. According to \cite{Bekki_2011}, such a process should be more 
important for lower-mass spiral galaxies ($\sim 1.2 \times 10^{10} M_\odot$), and for objects residing near the cluster core. Interestingly, \cite{Poggianti_2001} showed that faint red sequence S0 galaxies in the Coma cluster are on average younger than their bright counterparts and with evidence of recent star formation. Fig. 8 shows the co-added FORS2 spectra of XMM1229 red sequence galaxies grouped by morphological type. It can be noted that the $H\delta$ absorption line, a proxy for recent star formation, becomes stronger as one moves from elliptical to disc-dominated galaxies, suggesting that star formation terminated at more recent times in the latter and that these could be galaxies that just joined the red sequence at $z=0.98$.{\footnote{The S0 co-added spectrum shows a rather prominent absorption feature in the $H\delta$ region. We refrain from any conclusion from measurements performed on this feature, as the line appears specially narrow.}} A similar trend of the strength of the $H\delta$ with 
galaxy morphology was found by \cite{van_Dokkum_1998} and \cite{Tran_2007} in clusters at $z\sim0.3$ and $z\sim0.8$, respectively. Interestingly, \cite{Demarco_2010} showed similar trends between Balmer features and morphology for all cluster galaxies (not only red sequence members) in the galaxy cluster RX J0152.7-1357 at $z=0.84$ (see their Fig. 5). 

The MORPHS sample shows a higher fraction $F_{disc}$ of disc-dominated galaxies, compared to both XMM1229 and WINGS. We also note that, as in XMM1229, disc-dominated galaxies become more frequent at fainter magnitudes. \cite{Sanchez_Blazquez_2009} found that the fraction of red sequence early-type galaxies $F_{E+S0}$ decreases with cosmic time in the range $0.4 < z < 0.8$ in the EDisCS clusters, with a corresponding increase in the fraction of late-type galaxies. The depth of the MORPHS spectroscopic sample does not allow us to study the faint end of the red sequence and therefore to derive a reliable measurement of the disc fraction. However, \cite{Sanchez_Blazquez_2009} measured $F_{E+S0}=(57 \pm 9)$\% in MORPHS, which is consistent with the decreasing trend observed in their sample from 75\% to 55\% at $0.4 < z < 0.8$. These authors attributed the increase in $F_{disc}$ with cosmic time to the fact that, at the particular epochs spanned by EDisCS, many spiral galaxies joined the red 
sequence as they ceased to 
form stars. They 
also predicted a decrease of $F_{disc}$ at $z < 0.4$ due to morphological transformation of spiral galaxies into S0 galaxies. This agrees with what is observed in WINGS, where the fraction of disc-dominated galaxies is overall lower than the fraction of ellipticals and S0s and approximately constant with luminosity (see also \citealt{Valentinuzzi_2011} for analogous considerations). However, we stress here that the aim of the spectroscopic follow-up of MORPHS was the study of the Butcher-Oemler effect and therefore the target selection  was biased towards blue and spiral galaxies. Furthermore, we did not restrict the analysis of the spectroscopic MORPHS sample to within $0.54 \times R_{200}$, as this sample was already small. For this reason, although we used the correction factors of \cite{Poggianti_1999}, the $F_{disc}$ reported in Fig. 5 should be considered as an upper limit 

Studies of the morphology-density relation in samples of galaxy clusters up to $z=1$ have found that the overall fraction of early-type galaxies decreases with redshift (\citealt{Dressler_1997, Postman_2005}). However, \cite{Dressler_1997} and \cite{Postman_2005} conducted their studies on luminosity selected samples, while \cite{Holden_2007} found no significant decrease in the early-type fraction in a mass limited sample of cluster galaxies at $0.023 < z < 0.83$, suggesting that the evolution detected in magnitude limited samples must be significantly contributed by galaxies with stellar mass lower than $10^{10.6} M_\odot$, their adopted mass limit. Interestingly, \cite{Vulcani_2011} detected a decrease in $F_{E+S0}$ with redshift in their mass limited sample at $M > 10^{10.2} M_\odot$ and confirming it also with the \cite{Holden_2007} mass limit. A corollary to these results is that the increase in the late-type fraction with redshift corresponds to a decrease in the S0 fraction, with the fraction of 
elliptical galaxies not changing significantly. We note in Fig. 5 that the fraction of elliptical galaxies along the red sequence follows similar trends with magnitude in all the three samples considered.  \cite{Vulcani_2011} also found that the fraction of elliptical galaxies follows similar trends with stellar mass in the range $10.3 < \log(M/M_\odot) < 10.9$ in clusters up to $z \sim 0.8$ This suggests that elliptical and S0 galaxies in clusters follow different evolutionary paths, with the latter likely originating from the morphological transformation of passive spiral galaxies.

\cite{Mei_2009} investigated the morphological fractions along the red sequence of a composite sample consisting of the 8 clusters of the ACS Intermediate Redshift Cluster Survey. They found flat trends with magnitude for both the early and late-type morphological fractions in the inner regions of the clusters ($R_{cluster} < 0.6 R_{200}$). Unlike this work, they split the red sequence into bins of 1 magnitude each. We repeated the measurement of the morphological fractions using bins of one magnitude and we found that the trends observed in Fig. 5 became less pronounced and almost flat. In order to effectively compare with \cite{Mei_2009}, we will need to build a composite red sequence sample with all the HCS clusters, as the trends observed in XMM1229 could be peculiar only to this cluster.

\subsection{Structural Properties}

Figure 7 (right panel) shows the distribution of the S\'ersic index for red sequence galaxies in XMM1229, grouped according to their morphological type, while the median values of S\'ersic index and stellar mass are summarised in Table 6. The uncertainties quoted in this table correspond to the 68\% ($1\sigma$) width of the distribution of each quantity.\footnote{There is one only irregular galaxy on the red sequence, for which we find $n=0.90\pm0.09$ and $M_\star = 10^{9.8} M_\odot$.} We do not consider the behaviour of galaxy size in this paper as it is extensively discussed by Delaye et al. (2013) for the entire HCS sample.

In Fig. 7 we plot the S\'ersic index $n$, as a function of the $z_{850}$ magnitude (left panel) and stellar mass (central panel). From the central panel, it can be seen that there is some correlation between S\'ersic index and stellar mass, as it would be expected by \cite{Graham_2003}, although we find that it is rather weak ($\rho=0.3$). The comparison between the left and central panels of Fig. 7 shows that the stellar mass gives a clearer separation between early- and late-type galaxies, suggesting that even those disc-dominated galaxies that appear to be bright in the F850LP band are actually lower-mass objects. However, it is important to stress that the median fractional uncertainty on galaxy stellar mass is 24 \%, which is relatively high, and therefore this conclusion should be interpreted as an indication rather than an evidence of the build-up of the red sequence at low masses through quenching of star formation in spiral galaxies.

\begin{table}
 \centering
 \caption{Median S\'ersic indices and stellar masses, per morphological type, on the red sequence of XMM1229.}
 \resizebox{0.9\textwidth}{!}{\begin{minipage}{\textwidth}
   \begin{tabular}{|c|c|c|c|}
  \hline
  & Ellipticals & Bulge-dominated & Disc-dominated \\
  \hline
  \hline
  S\'ersic index (n) & $4 \pm 2$ & $2.6 \pm 1.1$ & $1.9 \pm 0.8$ \\
  \hline
  $\log{\frac{M_\star}{M_\odot}}$ & $10.5 \pm 0.5$ & $10.6 \pm 0.3$ & $10.2 \pm 0.2$\\
\hline
\end{tabular}
 \end{minipage}}
\end{table}

We note that the S\'ersic index of the elliptical galaxies spans a broad range of values. In particular, we note that some ellipticals have $n \sim 6$. These are mostly bright objects with $z_{850} < 22.0$, even though we obtain $n=6.6 \pm 0.6$ for the galaxy XMM1229\_353, which is a relatively faint object ($z_{850}=23.2$, $M_\star=10^{10.1} M_\odot$). From Fig. 6 (bottom-left panel) it can be seen that at the magnitudes of XMM1229\_353 $\Delta n = (n_{GALFIT}- n_{input})/n_{input}$ can be as large as 0.5, meaning that GALFIT can fit with a $n>4$ profile a galaxy that would be fitted by a de Vaucouleurs profile in an image with less sky noise than our F850LP image. Furthermore, as it is shown in Fig. 1 of \cite{Haeussler_2007}, the differences between $n=4$ and $n>4$ profiles are mainly in the outskirts of the galaxies, where the contamination from the sky is higher.

Interestingly, the value of $n$ for the BCG (XMM1229\_237) is: $n=6.4 \pm0.3$, which is consistent with results from local BCGs (see e.g. \citealt{Graham_1996}, although \citealt{Gonzalez_2005} found that a double de Vaucouleurs profile produces a better fit). We point out that the BCG resides in a crowded field (see Fig. 1) and therefore this result can be driven by contamination from intracluster light and close neighbours rather than from material that is associated to the BCG itself.

\begin{figure*}
  \includegraphics[width=0.7\textwidth]{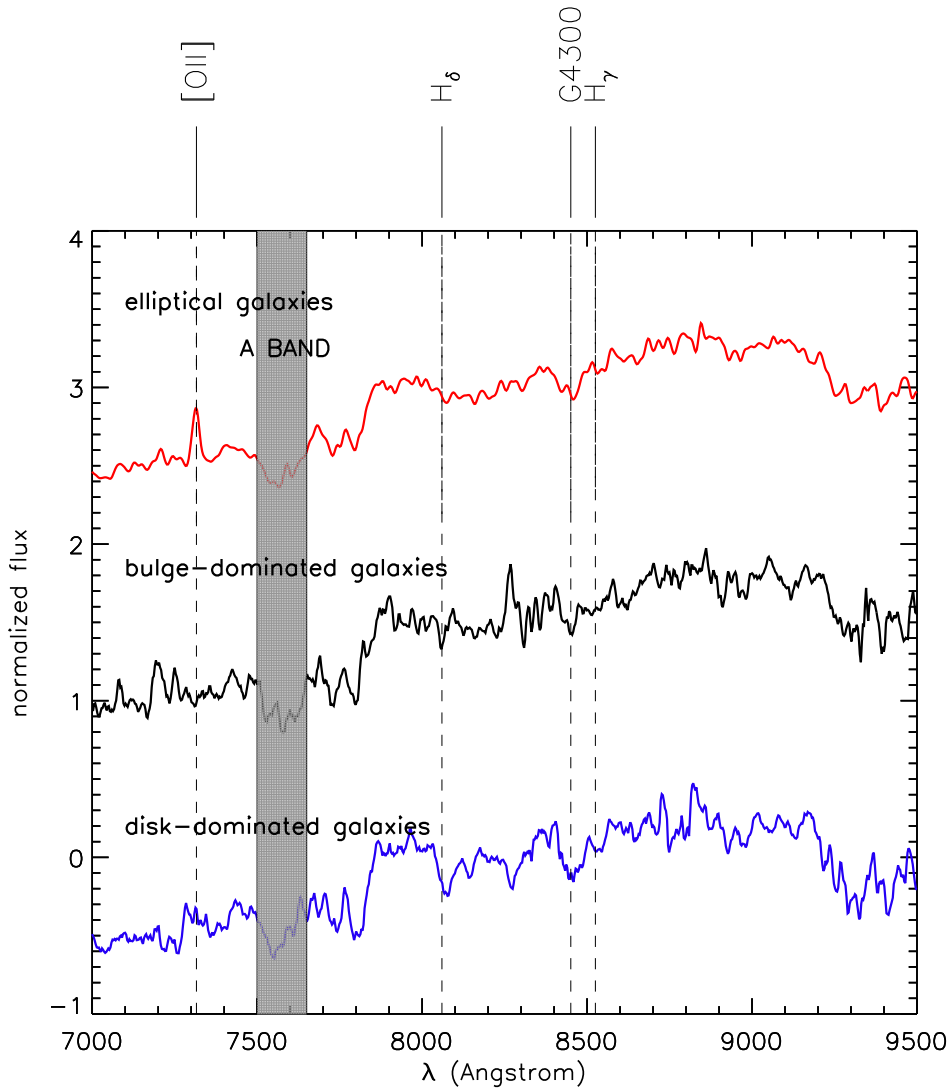}  
	\caption{Co-added FORS2 spectra of morphologically classified red sequence galaxies. The spectra are plotted in the observer frame and are shifted along the vertical axis for clarity. From top to bottom: elliptical, bulge-dominated, early disc-dominated galaxies. We highlight the $H\delta$, $H\gamma$ and G4300 features, and mask the $O_2$ atmospheric absorption (A band, $\lambda \sim 7604$ \AA). It can be seen that the $H\delta$ absorption is stronger in early disc-dominated galaxies. This suggests that these galaxies probably just joined the red sequence after cessation of star formation. It can be noted a remarkable [OII] emission feature in the elliptical spectrum, which is contributed by the galaxies XMM1229\_145 and XMM1229\_73. As reported by \protect\cite{Santos_2009}, these two spectra also show [OIII] in emission, which may indicate ongoing star formation.}
\end{figure*}

\subsection{The Red Sequence Slope and Scatter}

Prior to this work, the red sequence of XMM1229 was studied by \cite{Santos_2009} and \cite{Meyers_2012}, using the same F775W and F850LP images. In this section we describe the comparison between the results of the present and those works. 

The linear fit to the $(i_{775}-z_{850})$ versus $z_{850}$ red sequence produces a negative slope $b=-0.044 \pm 0.017$, consistent within the errors with \cite{Santos_2009} and \cite{Meyers_2012} ($b=-0.039 \pm 0.013$ and $b=-0.028$, respectively). Unlike \cite{Santos_2009}, we did not restrict our fit only to the spectroscopically confirmed members, as our sample of photometrically selected cluster members was 1 mag deeper. Delaye et al (2013) fitted the observed XMM1229 red sequence without any photometric or spectroscopic selection finding $b=-0.03$, a value closer to the \cite{Meyers_2012} result. \cite{Santos_2009} and \cite{Meyers_2012} found intrinsic scatters $\sigma_c = 0.039$ and $\sigma_c=0.066^{+0.003}_{-0.014}$, respectively, both consistent with our 
estimate $\sigma_c = 0.026 \pm 0.012$ within $1.1\sigma$ and $2.2\sigma$, respectively. 

The shallower slopes found by \cite{Meyers_2012} and Delaye et al. (2013) are a consequence of the fact that galaxy colours were estimated within one half-light radius in each galaxy in those works. As shown by \cite{Scodeggio_2001} and \cite{Bernardi_2003}, the steeper slope obtained with fixed apertures is to be attributed to the effect of residual colour gradients, which our large physical fixed aperture was not able to remove completely. However, as pointed out in both works, colours estimated within the galaxy half-light radius are also noisier and can produce large scatters, as demonstrated by the result of \cite{Meyers_2012}. Therefore, we think that the choice of a large fixed physical aperture, albeit unable to completely remove internal colour gradients, produces less noisy 
red sequences more suitable for evolutionary studies.

In agreement with most works in the recent literature on high redshift clusters, we find that the red sequence has a negative slope (\citealt{Lidman_2008, Mei_2009, Demarco_2010, Lemaux_2012, Snyder_2012}). Using the conversions from the observed $i_{775}$ and $z_{850}$ photometries to rest-frame B and V magnitudes derived in \S5.1, we find that our result is consistent with the median red sequence slope from \cite{Valentinuzzi_2011} $b = -0.042 \pm 0.007$. This supports the notion of little or no evolution of the slope of the red sequence since $z=1.5$ (see e.g. \citealt{Lidman_2008}). However, according to both hydrodynamical and N-body simulations \citep{Romeo_2008} and semianalytic models \citep{Menci_2008}, only based on hierarchical merging, the slope of the cluster red sequence should gradually evolve with redshift, flattening at $z \sim 0.7-1.0$, and then turning positive at higher redshifts. The results of the observations 
clearly indicate that these models were incomplete and other processes need to be taken into account to explain the evolution and the build-up of the red sequence. The simulations of \cite{Jimenez_2011}, based on a hybrid model constituted by a N-body cosmological simulation and a semi analytic model, produce red sequences with negative slopes for clusters at $z\sim1$, in better agreement with the observations. In these simulations, massive elliptical galaxies at the bright end of the red sequence constitute a primordial population which formed in the clusters and then evolved by accreting other galaxies at later epochs, thus shaping the red sequence as it is observed in the local universe.

\section{Summary and Conclusions}

We have presented a detailed analysis of the properties of red sequence members in the cluster XMMU J1229+0151 (XMM1229), at $z\sim0.98$. A study of the X-ray properties, as well as the analysis of the properties of the spectroscopically confirmed cluster members were presented by \cite{Santos_2009}. The availability of deep WFC3 images in four IR filters and HAWK-I Ks data for this field allowed us to fit synthetic spectral energy distributions and determine reliable photometric redshifts. We used the latter to estimate cluster membership, extending the analysis of the red sequence to about twice the galaxies considered in \cite{Santos_2009}, and down to one magnitude fainter than the limit of the FORS2 spectroscopic observations. 

Our estimates of the red sequence slope and scatter are consistent with the results of \cite{Santos_2009}, \cite{Meyers_2012} and Delaye et al. (2013), even though the latter two authors found a shallower red sequence and a larger scatter due to differences in the adopted strategy for aperture photometry. \textcolor{black}{The luminous-to-faint ratio measured on the red sequence of XMM1229 is higher than that measured in the WINGS composite sample, although the two estimates are consistent within the uncertainties}. The luminous-to-faint ratio in the cluster outskirts is consistent with no deficit of galaxies at the faint end of the red sequence.

After splitting into morphological classes, we found that the red sequence is predominantly populated by elliptical and S0 galaxies, whose fractions follow different trends with magnitude. As it is observed at lower redshifts, the bright end of the red sequence appears to be dominated by elliptical galaxies. Therefore, elliptical galaxies have been constituting the dominant class at the bright end of the red sequence since at least $z=1$. At intermediate luminosities, elliptical galaxies follow trends with magnitude which are similar to those observed in the MORPHS and WINGS composite samples. The faint end of the red sequence of XMM1229 is characterised by the increase in the fraction of disc-dominated galaxies. At similar absolute magnitudes, the WINGS red sequence is dominated by elliptical and S0 galaxies, suggesting that the latter may be the descendants of $z=1$ red spirals. We note that there is significant evolution of the bright end of the red sequence, resulting in a population of galaxies that are 
not observed at z=1. However, we caution against the fact that the WINGS red sequence extends to up to 2 mag brighter in the V band with respect to XMM1229 and 1.5 mag brighter with respect to MORPHS. We attribute this difference to the effect of colour gradients or biases in the WINGS optical and IR photometry, although the presence of young and/or metal-poor stellar populations in bright red sequence galaxies can result in bluer colours at the bright end of the red sequence, as observed in some of the WINGS clusters..

The method presented in this paper and devised to deal with archival data from different telescopes allows us to maximise the number of candidate cluster members on the red sequence using photometric redshifts or statistical background subtraction. It will be applied to the entire HCS sample to investigate the build-up of the red sequence at $0.8<z<1.5$ and to study the morphological evolution of its members.

\newpage

\section*{Acknowledgments}

We thank the anonymous referee for the helpful and constructive comments which contributed to improve this manuscript. We thank Joana Santos and Piero Rosati who supplied us the SofI J band image of XMM1229. We would like to thank Bianca Maria Poggianti and Alessia Moretti for providing us with the latest versions of the WINGS spectroscopic and morphological catalogues. We also thank Boris H\"{a}u{\ss}ler for providing the latest up to date scripts for galaxy image simulations. P.C. is the recipient of a Swinburne Chancellor Research Scholarship and an AAO PhD Scholarship. C.L. is the recipient of an Australian Research Council Future Fellowship (program number FT0992259). W.J.C. gratefully acknowledges the financial support of an Australian Research Council Discovery Project grant throughout the course of this work. R.D gratefully acknowledges the support provided by the BASAL Center for Astrophysics and Associated Technologies (CATA), and by FONDECYT grant N. 1130528. The data in this paper were based in 
part on observations obtained at the ESO Paranal Observatory (ESO programme 084.A-0214).

\appendix

\section{Visual Morphological Classification}

\textcolor{black}{The details of the morphological classification of red sequence members in XMM1229 are described in \S\S 4.4.1, 4.4.2 and Appendix B. In this appendix we present thumbnail images of the morphologically classified galaxies in each of the subsamples presented in Table 3. The postage stamp images were taken from the HST/ACS F850LP image of the XMM1229 field, which was used for morphological classification. Table 3 also lists positions and photometric and/or spectroscopic redshifts of all the objects. The images presented in the following figures belong, in the succession, to the following subsets: cluster centre, cluster outskirts, spectroscopically confirmed members that are excluded because they are not in the red sequence or because the photometry is inaccurate, spectroscopically confirmed members excluded from the analysis because they lie outside the area covered by the ACS images. For the last subsample the cutout images were taken from the HAWK-I Ks image. In each sample the objects 
are ordered according to their morphological type: elliptical, bulge-dominated, disc-dominated (early), disc-dominated (late), irregular. The classification for objects in the cluster outskirts corresponds to the output of galSVM. No visual classification was performed on this sample in the present work.} \textcolor{black}{The size of the ACS F850LP cutout images is 4.05$''$ on each side, corresponding to a physical size of 32.5 kpc at $z=0.98$. The size of the HAWK-I Ks cutout images is instead 8.1$''$ on each side, corresponding to a physical size of 65 kpc at $z=0.98$}

\begin{figure*}
	\includegraphics[trim=0.0cm 2.5cm 0.0cm 0.0cm, clip, width=16 cm]{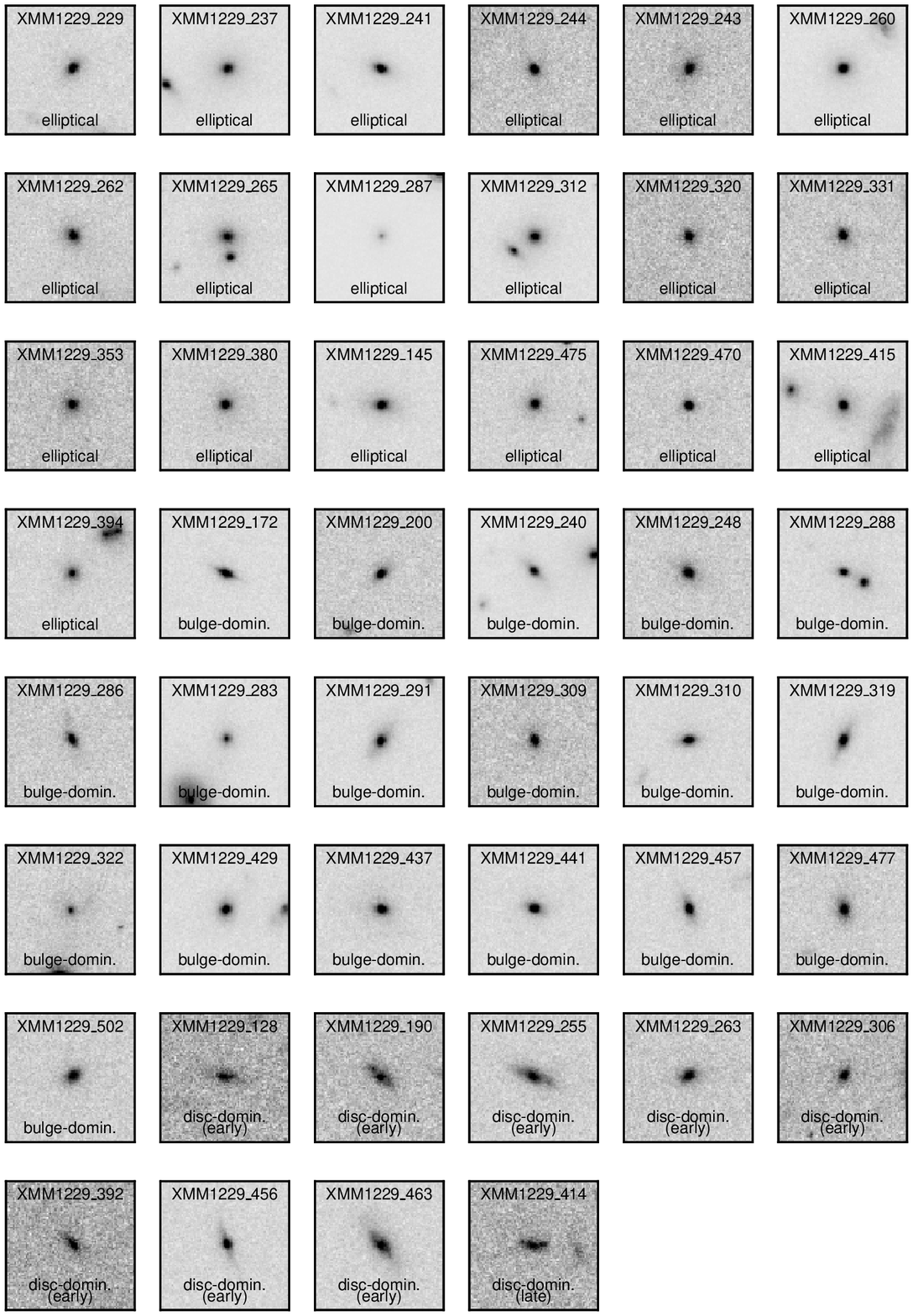}
	\caption{\textcolor{black}{Morphologically classified red sequence galaxies in the central region (i.e. within 0.6 Mpc from the cluster centre) . Position and redshift (photometric and/or spectroscopic) of each object are listed in Table 3. The image cutouts were taken from the HST/ACS F850LP image of the XMM1229 field used in morphological classification (see \S\S 4.4.1 and 4.4.2).}}
\end{figure*}

\begin{figure*}
	\includegraphics[trim=0.0cm 24.0cm 0.0cm 0.0cm, clip, width=16 cm]{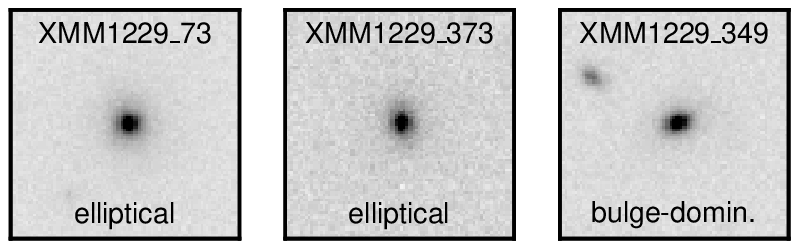}
	\caption{\textcolor{black}{Morphologically classified red sequence galaxies in the cluster outskirts (i.e. between 0.6 Mpc and 1.04 Mpc from the cluster centre). Only spectroscopically confirmed cluster members are shown, as in this work we do not determine the membership of individual galaxies for this subsample. Position and redshift of each object are listed in Table 3. The image cutouts were taken from the HST/ACS F850LP image of the XMM1229 field used in morphological classification (see \S\S 4.4.1 and 4.4.2). The morphological classification of the objects in this subsample corresponds to the outcome of galSVM.}}
\end{figure*}

\begin{figure*}
	\includegraphics[trim=0.0cm 24.0cm 0.0cm 0.0cm, clip, width=16 cm]{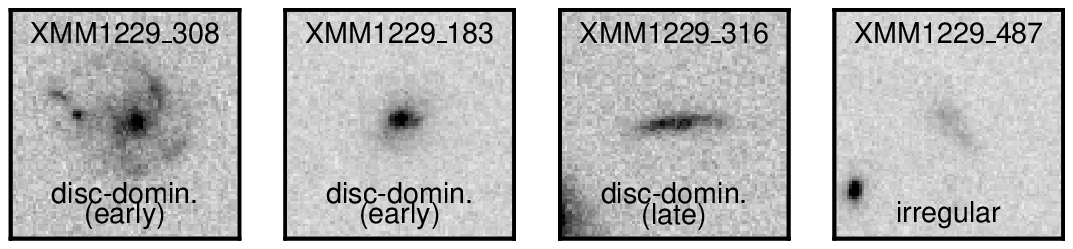}
	\caption{\textcolor{black}{Spectroscopically confirmed XMM1229 members excluded from the analysis in this paper. XMM1229\_183, XMM1229\_308, XMM1229\_487 are blue cloud galaxies, while XMM1229\_316 was excluded because SExtractor was unable to return a reliable estimate of the aperture magnitude (SExtractor {\ttfamily{FLAGS}}=16). The image cutouts were taken from the HST/ACS F850LP image of the XMM1229 field. For this subsample the visual morphological classification was performed by P. C. .}}
\end{figure*}

\begin{figure*}
	\includegraphics[trim=0.0cm 24.0cm 0.0cm 0.0cm, clip, width=16 cm]{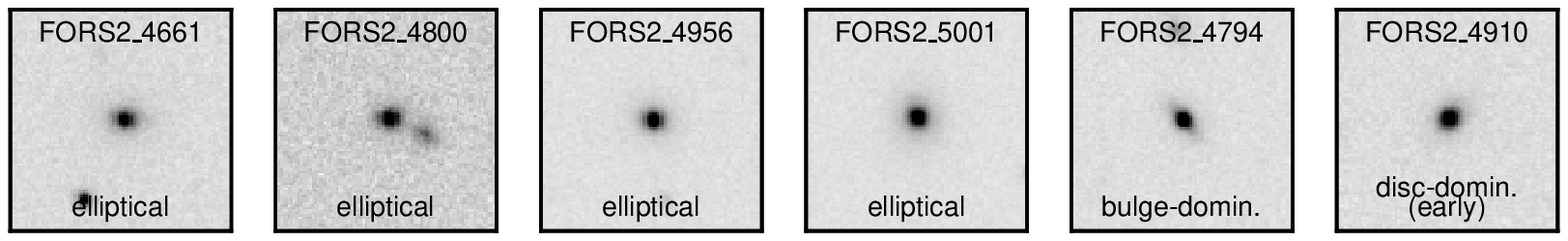}
	\caption{\textcolor{black}{Spectroscopically confirmed XMM1229 members excluded from the analysis in this paper because falling outside the ACS field. The classification was performed visually by P. C. on image cutouts taken from the HAWK-I Ks band image of the XMM1229 field and shown in this figure.}}
\end{figure*}

\section{SVM-based Galaxy Morphology}

{\ttfamily galSVM} is based on Support Vector Machines (SVM), a particular family of 
machine-learning algorithms that, given a training sample of objects with known classification, 
fits a hyperplane in the space of the measured parameters to separate between two particular classes. 
In the case of galaxy morphology, the parameter space is defined by a certain number of 
morphological coefficients measured on the images. For the XMM1229 field, we measured and used seven non-parametric morphological coefficients: 
concentration (as defined in \citealt{Conselice_2000} and \citealt{Abraham_1996}), asymmetry, Gini coefficient \citep{Abraham_2003}, $M_{20}$ (\citealt{Lotz_2004}), smoothness \citep{Conselice_2003} and ellipticity (as measured by SExtractor). We refer to \cite{Huertas_2008} for a more detailed explanation of these quantities. The training set used in this classification was generated from a sample of $\sim$ 14,000 visually classified galaxies from g-band images of the SDSS DR7 \citep{Nair_2010}. In order to reproduce as much as possible the conditions of high redshift galaxies, the images were degraded according to the $z_{phot}$ and F850LP {\ttfamily MAG\_AUTO} distributions in the XMM1229 field. The advantage of this technique is that it takes into account the effect of both redshift and flux on the detection of morphological features and at the end of the degradation process one has a sample of galaxies of known morphologies as they would appear if they were at 
the redshift of the galaxies in the XMM1229 F850LP image.

SVM are binary classifiers and, in order to reproduce a morphological scheme similar to the one adopted in the visual classification, {\ttfamily galSVM} needed to be run more than once. In the first run, {\ttfamily galSVM} used the entire training sample to classify galaxies into early- and late-type. Then, two subsequent runs, using only early- or late-type galaxies in the training sample, allowed the software to split the early-type broad class into elliptical and S0 galaxies, and the late-type broad class into early discs and late discs plus irregular galaxies. In order to estimate the uncertainty on the classification, {\ttfamily galSVM} implements a scheme of posterior probabilities which quantify the likelihood of the classes assigned by the SVM classifier. The morphological type is the one that maximises the composite conditional probability:
\begin{equation}
 P(T) = P(BT)P(T|BT)
\end{equation}
where T is the morphological type and BT is the broad morphological class, that can be either early- (ETG) or late-type (LTG). 
We assigned morphological classes to the galaxies classified by {\ttfamily galSVM} following the subsequent criterion:
\begin{itemize}
 \item {\bf{ellipticals}}: galaxies with $P(ETG)>0.5$ and $P(E | ETG)>0.5$;\\
 \item {\bf{bulge dominated}}: galaxies with $P(ETG)>0.5$ and $P(E | ETG) \leq 0.5$;\\
 \item {\bf{early disc-dominated}}: galaxies with $P(ETG) \leq 0.5$ and $P(EDD | LTG) > 0.5$;\\
 \item {\bf{late disc-dominated plus irregulars}}: galaxies with $P(ETG) \leq 0.5$ and $P(EDD | LTG) \leq 0.5$;\\
 \end{itemize}
With this definition, each galaxy has still a finite but lower probability of belonging to a different morphological class to that assigned, and this quantifies the uncertainty on the classification.

Figure B1 shows the distributions of concentration, asymmetry, Gini coefficient and $M_{20}$ for red sequence galaxies in the centre of XMM1229 classified as described in \S 4.4.1. Types were assigned to each galaxy following a majority rule in which the morphological class was defined as the mode of the four independent classifications. As expected by e.g.\ \cite{Lotz_2004}, disc-dominated galaxies have lower values of concentration and Gini coefficient. However, the values of $M_{20}$ appear still comparable with those of elliptical and S0 galaxies. This can be attributed to the fading of the spiral arms in gas-poor spiral galaxies.

\begin{figure*}
	\includegraphics[width=16 cm]{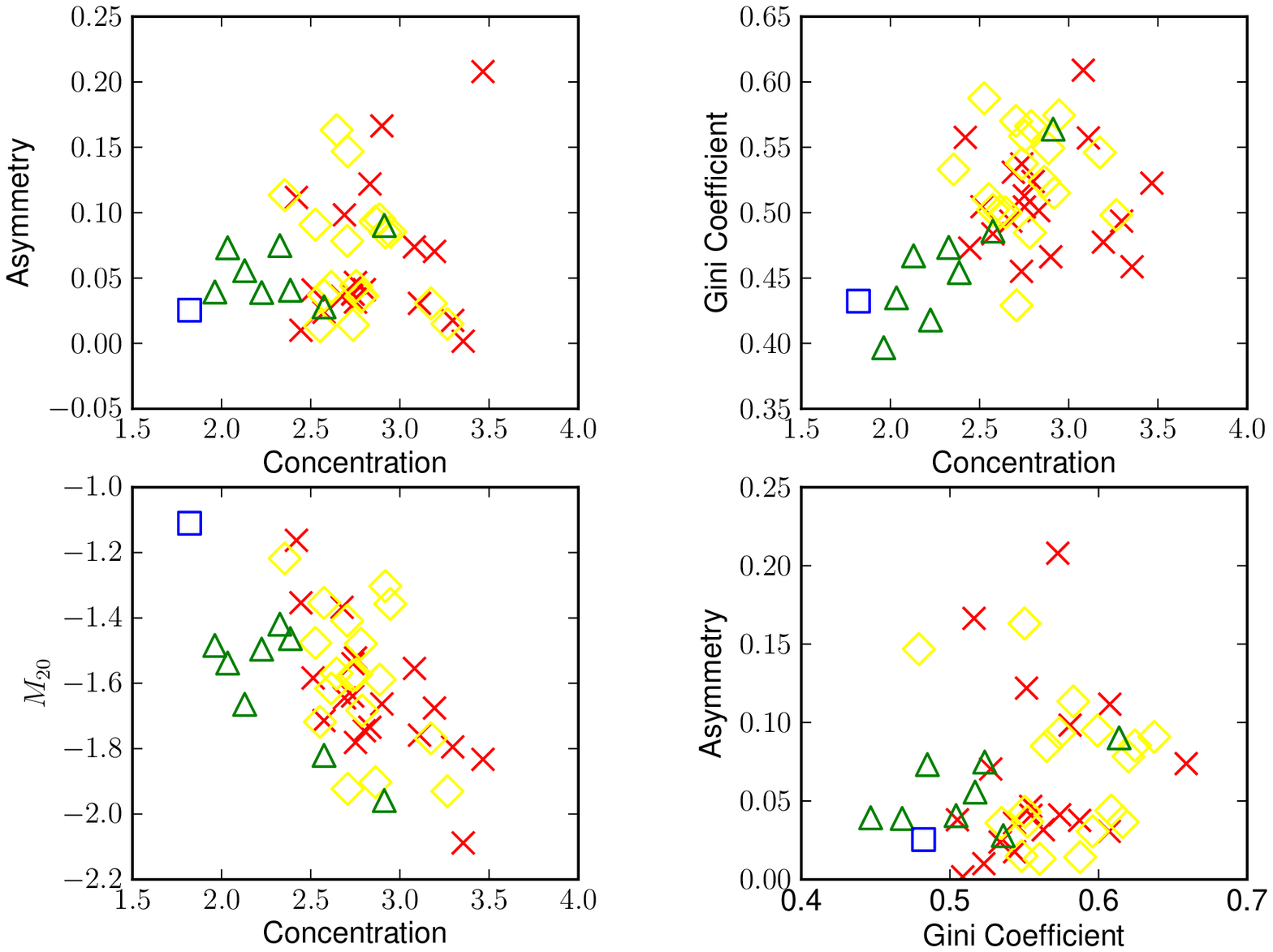}
	\caption{Morphological parameters for red sequence galaxies in the centre of XMM1229 ($R_{cluster}<0.54 \times R_{200}$). Symbols and colours are the same used in Fig. 5. As expected, disc-dominated galaxies tend to have smaller values of Concentration and Gini coefficient. All morphological types show comparable values of $M_{20}$.}
\end{figure*}

\bsp

\label{lastpage}

\end{document}